\documentclass[fleqn,usenatbib]{mnras}

\usepackage{newtxtext,newtxmath}
\usepackage{soul}
\usepackage{amsmath}
\usepackage{graphicx}
\usepackage{booktabs}
\usepackage{tabularx}
\usepackage{caption}
\usepackage{float}
\usepackage{microtype}
\usepackage{placeins}
\usepackage[usenames]{xcolor}
\usepackage{color}
\usepackage{soul}

\usepackage[T1]{fontenc}

\DeclareRobustCommand{\VAN}[3]{#2}
\let\VANthebibliography\thebibliography
\def\thebibliography{\DeclareRobustCommand{\VAN}[3]{##3}\VANthebibliography}

\usepackage{graphicx}	
\usepackage{amsmath}	
\usepackage{float}
\usepackage{orcidlink}

\title[Companion induced shadows in discs]{Modelling shadows in scattered light observations as signals from companions in protoplanetary discs}

\author[]{
Deniz Akansoy$^{\orcidlink{0009-0006-3784-3177}1, 2}$\thanks{E-mail: da619@cam.ac.uk}, 
Helen Petrou$^{\orcidlink{0009-0008-2388-9291}2}$, 
Giulia Ballabio$^{\orcidlink{0000-0002-4687-2133}2}$\thanks{E-mail: g.ballabio@imperial.ac.uk}
and Anna Penzlin$^{\orcidlink{0000-0002-8873-6826}2,3}$
\\
$^{1}$Department of Physics, University of Cambridge, Cambridge CB3 0HE, UK \\
$^{2}$Imperial Astrophysics, Imperial College London, Blackett Laboratory, Prince Consort Road, London SW7 2AZ, UK \\
$^{3}$Ludwig-Maximilians-Universit{\"a}t M{\"u}nchen, Universit{\"a}ts-Sternwarte, Scheinerstr.~1, 81679 M{\"u}nchen, Germany\\
}

\date{Accepted XXX. Received YYY; in original form ZZZ}

\pubyear{2024}

\begin{document}
\label{firstpage}
\pagerange{\pageref{firstpage}--\pageref{lastpage}}
\maketitle

\begin{abstract}
Over the past decade, SPHERE scattered light observations of protoplanetary discs have revealed previously unseen features with unprecedented resolution. One such feature are radial streaks of reduced brightness that are commonly interpreted as shadows. A possible cause for these shadows is an embedded companion within the disc. In this work, we use 3D radiative transfer simulations with \texttt{RADMC-3D} to investigate the shadowing effects of embedded companions across a range of orbital distances (5-30 au) and companion masses (0.5-30 $M_\text{J}$). We model $0.1~\mu$m dust grains, which are well-coupled to the gas, to produce synthetic scattered light images of the disc. Companions with masses equal to or greater than 14 Jupiter masses consistently cast detectable shadows throughout the disc. We hence derive an empirical solution to describe the width and depth of the shadow as functions of companion mass and location. This scaling suggests that shadow features observed in scattered light images could serve as reliable indicators of companion mass and position, providing an indirect method for identifying and characterising otherwise challenging-to-detect objects within these discs. Additionally, our analysis reveals that companion shadows influence the disc thermal structure, with notable cooling effects that could impact disc chemistry and the dynamics of planet formation. 
\end{abstract}




\section{Introduction} 
Studying protoplanetary discs (PPDs) can reveal vital information regarding the origins and evolution of planetary systems. Within the last decade, observations of PPDs have evolved from simple Spectral Energy Distributions to highly resolved images of the continuum and scattered light emission. In recent years, by utilising the capabilities of Spectro-Polarimetric High-contrast Exoplanet REsearch (VLT/SPHERE; \citep{buezit_2019}), programmes like  GTO and DESTINYS \citep{2024DESTINYS_Cha,2024DESTINYS_Ori,2024DESTINYS_Tau} have comprehensively surveyed PPDs across entire star-forming regions in near-infrared scattered light. This has led to the discovery of various disc substructures, with azimuthally asymmetric shadows being among the most interesting.
\begin{figure}
    \centering
    \includegraphics[width=0.75\columnwidth]{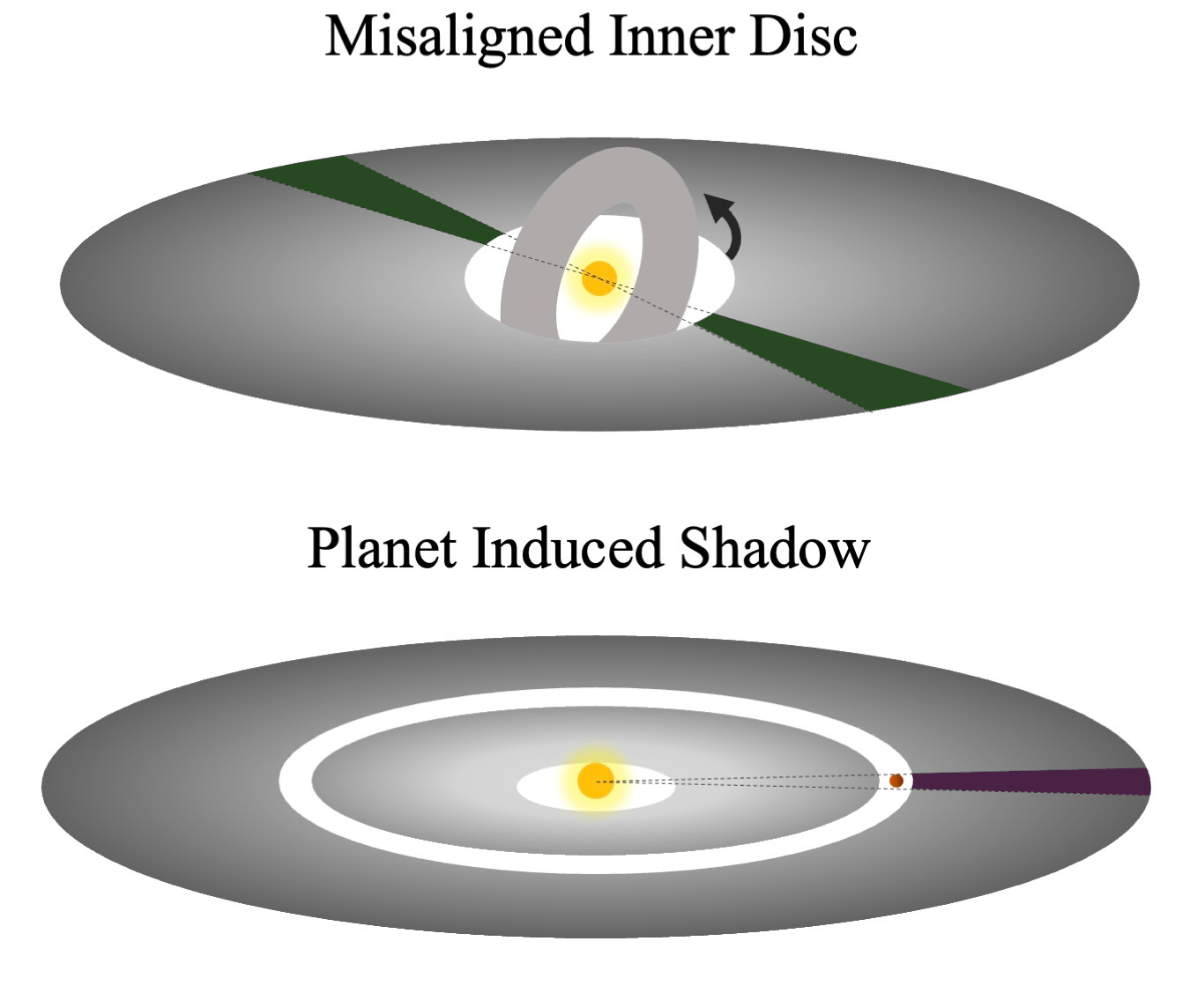}
    \caption{Possible disc configurations causing shadows. \textit{Left:} Symmetric shadows caused by an extremely misaligned disc. \textit{Right:} Shadow caused by a companion exceeding the local disc scale height and directly blocking light. }
    \label{fig:shadow origins}
\end{figure}

\begin{figure}
    \centering
    \includegraphics[width=0.8\columnwidth]{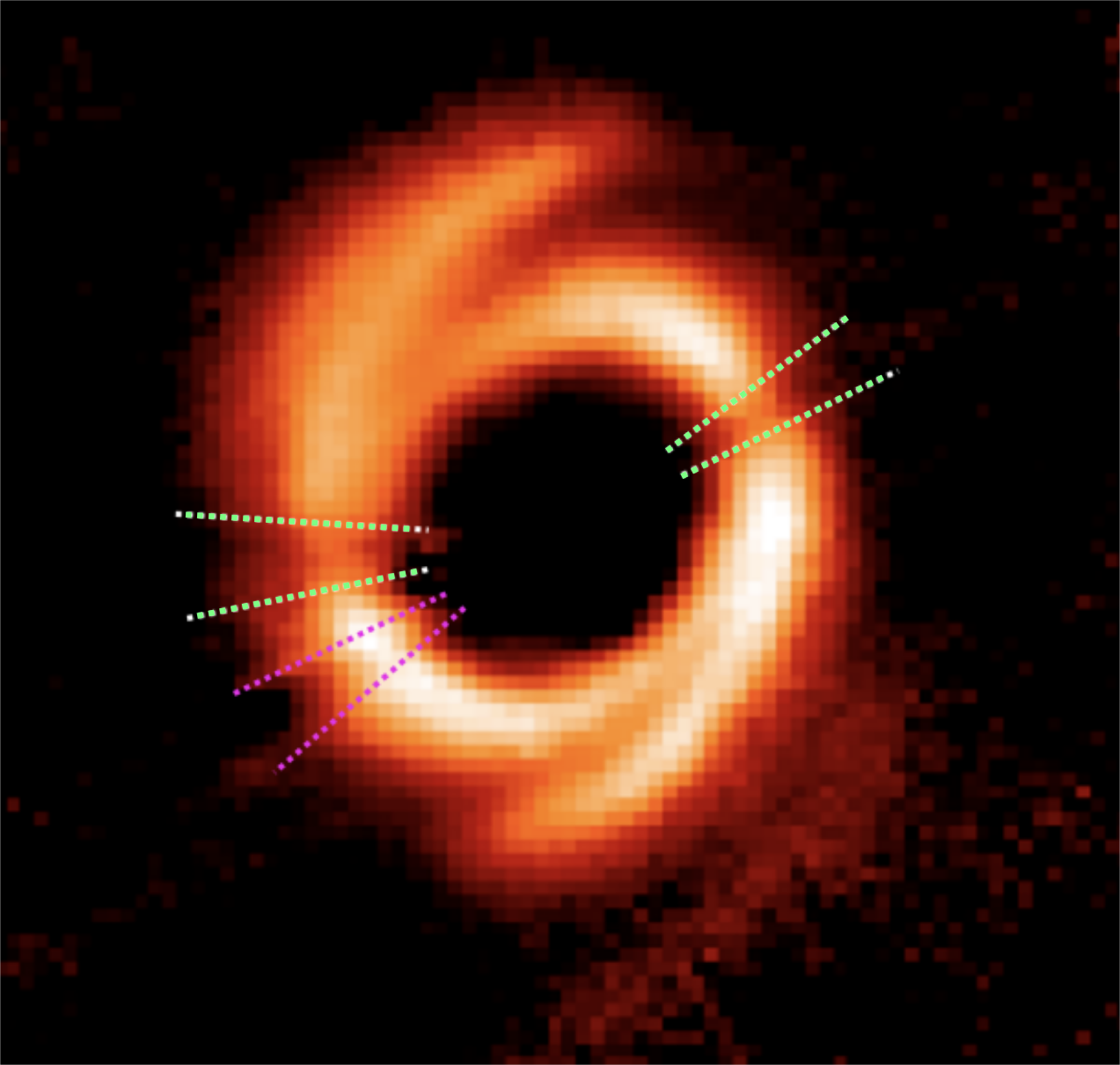}
    \caption{A near-infrared image of the protoplanetary disc HD100453, observed by VLT/SPHERE. Bright green dotted lanes mark the location of shadows caused by the misalignment of the inner disc, while the pink dotted line marks the location of the potential third shadow thought to be caused by a planet embedded within the disc. Image adapted from \protect\cite{2017Benisty} \protect\footnotemark[1].}
    \label{fig:HD100453}
\end{figure}

Narrow, shadowed lines in several complex discs, including GG Tau \citep{2020Keppler}, HD 135344B \citep{2016Stolker} and HD 100453 \citep{2017Benisty}, have been revealed through these studies, with evidence of moving shadow features observed in HD 169142 \citep{2020Bertrang, 2022Poblete}. While symmetric shadows are often linked to misaligned inner discs, isolated shadows may suggest the presence of a faint companion enveloped by accreting material. Disc arrangements for both scenarios are illustrated in Figure \ref{fig:shadow origins}. Theories pointing at irregular clumps or dust clouds elevated within the disc have been put forward to explain observations of isolated shadows \citep{Rich_2019}. In line with this, it is theorised that protoplanets or companions may be able to scatter enough light to cast their own shadows \citep{Montesinos_2021}. Three-dimensional simulations of accretion onto planets suggest that infall from above the PPD can enhance the shadowing effect \citep{2019Fung, 2020Mattaus, 2024Lega}, especially when the material fills a significant portion of the companion's Hill sphere. Depending on the companion's mass, this accumulation can exceed the local disc height, casting distinct shadow streaks in scattered light. An example of a disc with shadows, HD100453, can be seen in Figure \ref{fig:HD100453}. The two shadow slits indicated by green dotted lines are thought to result from a $\sim$72$^{\circ}$ misalignment of the inner disc, casting shadows on the outer disc \citep{2019Zhu}. A speculated third shadow slit, shown with pink dotted lines, may be caused by a planet embedded within the disc casting its own shadow. 

\footnotetext[1]{\textcopyright~Reproduced and adapted with permission from \cite{2017Benisty}.}

Despite the increasing numbers of observed shadows in PPDs, our theoretical understanding of the shadow profile caused by planets and companions remains incomplete. This paper aims to bridge that gap by employing \texttt{RADMC-3D} radiative transfer simulations \citep{radmc} to model shadow-casting embedded objects within discs. The goal is to develop a parameterisable solution for the strength and size of the shadowed region, depending on the companion's properties. Verifying this could unlock opportunities for detecting newly-forming companions that are otherwise undetectable through other imaging techniques. Additionally, concentric gaps observed in scattered light images of PPDs are commonly attributed to planets \citep[e.g.][]{2021Kanagawa}, but they can also arise from other processes like pressure feature due to ice-lines \citep{2016Okuzumi}, variations in the radial gas drift e.g. at the dead zone edge \citep{2016Pinilla}, or thermal-dynamic disc interaction caused for example by an inner misaligned disc \citep{2024Ziampras}. Studying shadows in scattered light can offer further confirmation of a companion's role in creating these gap features.

In this work, we simulate the effects of an embedded companion in a disc using radiative transfer simulations and find an analytical solution to describe the shadows produced. Section \ref{sec:method} introduces the geometric model of the disc-planet system used for the simulations and the post-processing work done. Section \ref{sec:results} presents the outcomes of the simulations and a characterisation of the shadows, as well as an analysis on the temperature structure in the disc. Section \ref{sec:discussion} serves as a discussion on our disc model and analytical forms found for the shadow characteristics, and discusses the implications of the temperature structure of the disc. Section \ref{sec:conclusions} states our main findings.

\section{MODEL \& METHODS} \label{sec:method}

To determine how light is scattered, the radiative transfer calculations require a 3D model of the dust distribution within the protoplanetary disc. In order to model the disc, we first create a 3D grid in spherical coordinates and assign gas and dust density values to each cell. The grid consists of 300 radial cells (2 – 100 au) that are evenly spaced in log-scale, with an additional grid refinement near the inner edge, increasing the total number of radial cells to 336. The grid also includes 100 polar cells within $\pi/4 < \theta < \pi/2$ assuming midplane symmetry, and 300 azimuthal cells spanning $0 < \varphi < 2\pi$. Hence the resolution $N_r \times N_\phi \times N_\theta $ is $336\times 300 \times 200$. Our analytic disc model can be broadly divided in three steps: 
\renewcommand{\labelenumi}{\alph{enumi})}
\begin{enumerate}
\item{Building a flaring disc model}
\item{Implementing the gap that would be carved out by a companion}
\item{Embedding the companion}
\end{enumerate}

\noindent This model is then used as an input for radiative transfer simulations in order to obtain the final scattered light images. With this choice of number of cells, the grid is sufficiently refined in both the radial and azimuthal directions to ensure that companion-induced structures, including any cast shadows, are resolvable (e.g. the hill sphere of the $14~M_\mathrm{J}$ companion is covered by 11 cells with this resolution).

\subsection{Disc Model} \label{sec:gas_dist}
We model the gas and dust density within the disc analytically, assuming that the disc is non self-gravitating ($M_* \gg M_\mathrm{disc}$). In cylindrical coordinates, the density distribution, $\rho(r, z)$, for a disc in vertical hydrostatic equilibrium can be written as,

\begin{equation}
    \rho (r, z) = \frac{1}{\sqrt{2\pi}}\frac{\Sigma(r)}{H(r)}\exp\left[{-\frac{1}{2}{\left(\frac{z}{H(r)}\right)}^2}\right],
    \label{eqn: vertical distribution}
\end{equation}

\noindent where the density varies both radially and vertically, while assumed to be azimuthally symmetric. We choose to model the surface density, $\Sigma'(r)$, as a power law of the form
\begin{equation}
    \Sigma'(r) = \Sigma_0 \left(\frac{r}{\text{au}}\right)^{-1}
    \label{eqn: surface density}
\end{equation}

\noindent where $\Sigma_0$ is the surface density at 1 au. We use a single dust grain size of 0.1 $\mu$m, thereby, the Stokes number of the dust remains below $10^{-4}$, and is well coupled to the gas. We model our disc to have a dust surface density of 0.1 g$\,$cm$^{-2}$ of $0.1~\mu$m grains at 1~au, with a scale height, $H(r)$, defined using the relation,
\begin{equation}
    H(r) = c_\text{s} \cdot \Omega_\text{K}^{-1} = \sqrt{\frac{k_\text{B} T_\text{mid}(r)}{2.3\,m_\text{p}}}\cdot \sqrt{\frac{r^{3}}{GM_*}}
\end{equation}
\noindent where the sound speed at the disc midplane, $c_\text{s}$, depends on the Boltzmann constant, $k_\text{B}$, the proton mass, $M_\text{p}$, and the midplane temperature, $T_\text{mid}$. The Keplerian angular frequency, $\Omega_\text{K}$, depends on radial distance, $r$, the gravitational constant, $G$, and the stellar mass, $M_*$. The midplane temperature, $T_\text{mid}$, is calculated analytically for a flaring disc, as presented by \cite{Truelove_1997}, with the relation,
\begin{equation}
    T_\text{mid} (r) = \varphi^{1/4} \left(\frac{R_*}{r}\right)^{1/2} T_* \propto r^{-1/2}.
\end{equation}
The flaring angle, $\varphi$, is kept constant at a value of 0.25, $T_*$ is the blackbody temperature of the star, and $R_*$ is the star radius. The radial dependence of $T_\text{mid} \propto r^{-1/2}$ results in the pressure scale height being governed by $H(r)\propto r^{5/4}$. Consequently, the disc aspect ratio, $h(r) ={H(r)}/{r} = r^{\eta}$, has a positive flaring index of $\eta = 0.25$ and so the disc is said to be flared.

\begin{figure}
    \centering
    \includegraphics[width=\columnwidth]{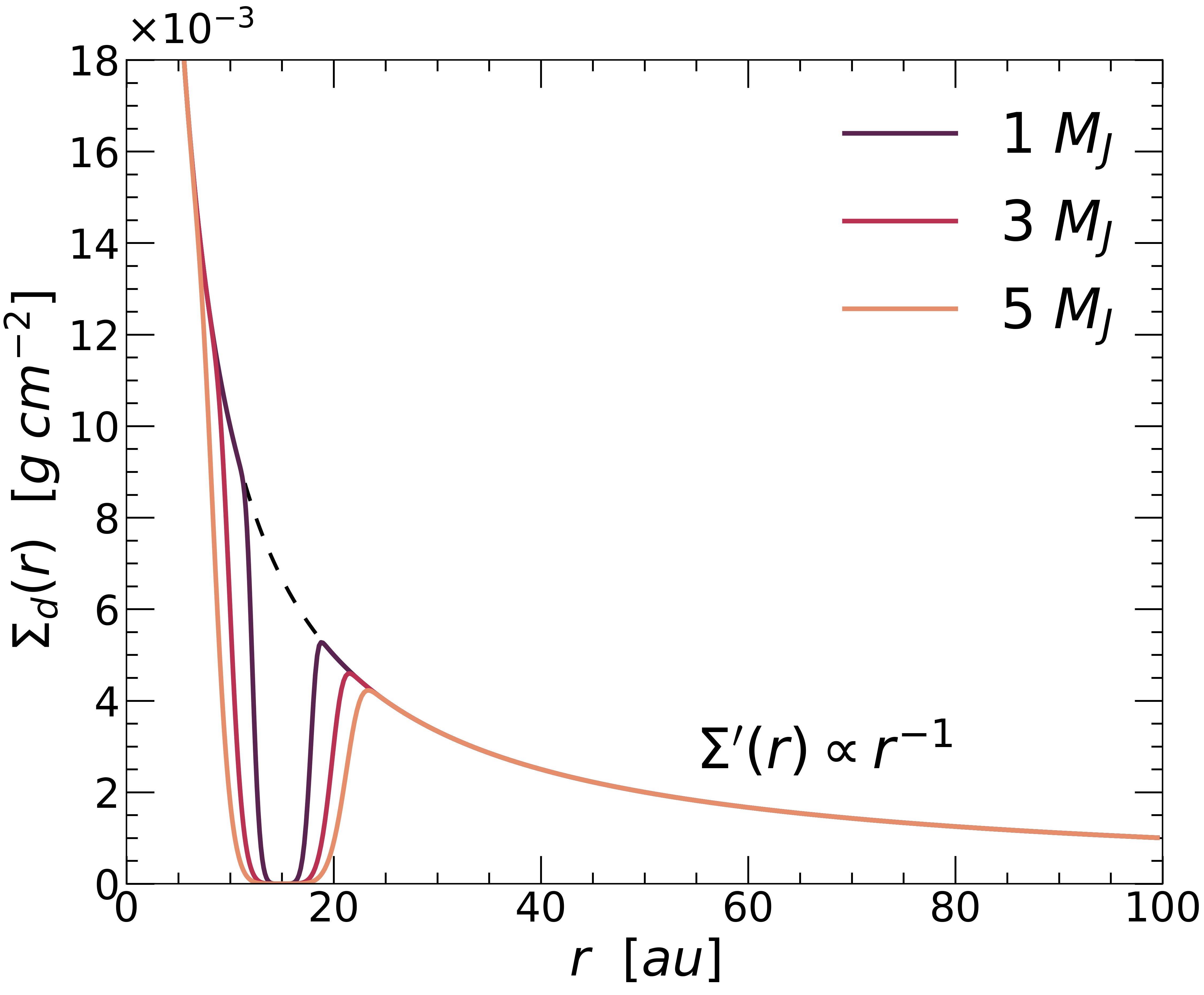}
    \caption{Dust surface density plotted against radius, for a disc with a gap centered at $r_\text{p} =$ 15 au. The dotted line represents the disc with no gap, i.e., $\Sigma(r)=\Sigma(r)'$.}
    \label{fig:gap widths}
\end{figure}

In order to implement the gap a companion would carve within the disc, we first determine the gap width, $\Delta_\mathrm{gap}$, using the empirical relation presented by \cite{kanagawa_2016},  
\begin{equation}
    \frac{\Delta_\mathrm{gap}}{r_\text{p}} = 0.41 \left(\frac{M_\text{p}}{M_*}\right)^{1/2}\left(\frac{H_\text{p}}{r_\text{p}}\right)^{-3/4}\alpha^{-1/4},
    \label{eqn:kanagawa}
\end{equation}
where $r_\text{p}$ is the orbital separation of the companion from the host star, $M_\text{p}/M_*$ is the fraction of the companion mass against the stellar mass, $H_\text{p}/r_\text{p}$ is the disc aspect ratio at the companion's orbital distance, and $\alpha$ is the Shakura-Sunyaev viscosity parameter \citep{Shakura_sunyaev1973}. This relationship shows that the width of the gap will increase with increasing mass (Figure \ref{fig:gap widths}) and orbital distance.

After calculating the gap width, we modify the surface density profile, $\Sigma(r)$, as
\begin{equation}
    \Sigma(r) = \Sigma'(r) \left[ 1 -  \exp{ \left( -\frac{1}{2} \left(\frac{r - r_\text{p}}{0.4 \; \Delta_\text{gap}}\right)^{6} \right) } \right]
    \label{eqn:mod_surf_dens}
\end{equation}
where $\Sigma'(r)$ is the initial surface density from Equation \ref{eqn: surface density}. The term in the square brackets is a bell-shaped gap cut-out centered at the planet location, $r_\text{p}$, with its width governed by $\Delta_{\text{gap}}$. We have assumed that the surface density falls to 0 at $r_\text{p}$. The power of 6 in the exponent was chosen so that the edges of the gap were steeper than a Gaussian, to match the gap shapes opened in hydrodynamic simulations \citep{2006Crida}. 

We then embed a companion into our disc model at $r = r_\text{p}$, $\theta = \pi/2$, $\varphi = 0$ in spherical coordinates, assuming it forms in the midplane $(\theta = \pi/2)$. Instead of determining a companion radius, we choose to model the companion and nearby gravitationally bound material as a spherical distribution of matter extending up to the companion's Hill radius, $r_\text{H}$. This is calculated using
\begin{equation}
    \frac{r_\text{H}}{r_\text{p}} \approx \sqrt[3]{\frac{M_\text{p}}{3\:\left(M_* + M_\text{p} \right)}} \approx  \sqrt[3]{\frac{M_\text{p}}{3\:M_*}}
\end{equation}
which is an approximation under the assumptions that $r_\text{p} \gg r_\text{H}$ and $M_* \gg M_\text{p}$. This equation also treats the companion and star as point masses, ignoring the disc's gravitational influence (assuming a non-self-gravitating disc, $M_* \gg M_\text{disc}$). After calculating the Hill radius, the density distribution within the companion sphere is modelled as a 3D Gaussian of the form
\begin{equation}
    \rho(r') = \Sigma_\text{d}(r_\text{p}) \exp{\left[ -\frac{1}{2} \left( \frac{r'}{r_\text{H}/3} \right)^{2} \right]}
\end{equation}
where $r' = |\vec{r_\text{p}} - \vec{r}|$ is the radial distance from the centre of the companion. This avoids defining a solid surface for the protoplanet and accounts for matter flowing in and out of the companion region. The factor of 1/3 ensures that the model places 3$\sigma$ (>99\%) of the circumplanetary material within the Hill sphere, representing the optical effect of the companion-bound region.
Assuming a nominal gas surface density of $1000\ \mathrm{g\ cm^{-2}}$ at 1 au and a micrometre-sized dust mass fraction of $10^{-4}$, the dust at 100 au has Stokes numbers smaller than $10^{-4}$. As a simplifying assumption in the model, we therefore treat the dust as perfectly coupled to the prescribed gas profile.

We assume the dust to solely be amorphous olivine, composed of 50\% magnesium and 50\% iron. The numerical values for material density and grain size are displayed in Table \ref{tab:paramerers}. To streamline computational resources, our investigation is directed towards parameters anticipated to have a significant impact on the casting of shadows, namely the companion mass (in Jupiter masses, $M_\text{J}$) and its orbital distance. Table \ref{tab:paramerers} outlines the range of values investigated for these parameters, alongside a summary of the fixed model parameters. Some of the orbital distances selected are analogous to those within the solar system: Jupiter at 5 au, Uranus at approximately 20 au, and Neptune at 30 au.
\begin{table}
    \centering
    \caption{Input parameters of the analytical model. Companion masses range from 0.5 $M_\text{J}$ to 30 $M_\text{J}$, with specific values of 0.5 $M_\text{J}$, 1 $M_\text{J}$, 5 $M_\text{J}$, 10 $M_\text{J}$, and from 14 $M_\text{J}$ to 30 $M_\text{J}$ in 2 $M_\text{J}$ intervals. Companion distances range from 5 au to 30 au in 5 au intervals, with an additional distance of 7.5 au.}
    \begin{tabularx}{\columnwidth}{l>{\hspace{0.4cm}}c>{\hspace{0.9cm}}c}
        \toprule
        \textbf{Parameter} & \textbf{Symbol} & \textbf{Value(s)} \\
        \midrule
        \multicolumn{3}{c}{\textbf{Fixed Parameters}} \\
        \midrule
        Star mass & $M_\ast$ & $M_\odot$ \\
        Star radius & $R_\ast$ & $R_\odot$ \\
        Star temperature & $T_\ast$ & 5780 K \\
        Disc inner radius & $R_{\text{in}}$ & 2 au \\
        Disc outer radius & $R_{\text{out}}$ & 100 au \\
        Flaring index & $\eta$ & 0.25 \\
        Viscosity Alpha & $\alpha$ & $4 \times 10^{-4}$ \\
        Dust grain size & $a$ & 0.1 $\mu$m \\
        Dust material density & $\rho_\text{d}$ & 3.71 g cm$^{-3}$ \\
        Dust surface density at 1 au & $\Sigma_{0,\text{d}}$ & 0.1 g cm$^{-2}$ \\
        \midrule
        \multicolumn{3}{c}{\textbf{Free Parameters}} \\
        \midrule
        Planet (Companion) mass & $M_\text{p}$ & 0.5 - 30 $M_\text{J}$ \\
        Planet (Companion) distance & $r_\text{p}$ & 5 - 30 au \\
        \bottomrule
    \end{tabularx}
    \label{tab:paramerers}
\end{table}

\subsection{Radiative Transfer}

In order to simulate radiative transfer within our analytic model, we use the software package \texttt{RADMC-3D} \citep{radmc} which calculates directional scattering and absorption of photon packages in an irradiated dust environment. Since shadows are mainly seen in scattered light images, we first focus on the $1~\mu$m scattered light signal from micron-sized dust particles suspended on the surface of the disc.
The mass-weighted absorption and scattering opacities are obtained via Bohren and Huffman's method for calculating Mie-scattering and absorption, presented in \cite{bohren1998} (method provided in the \texttt{RADMC-3D} package). The optical constants for olivine used as dust in our model are obtained from the measurements presented in \cite{1994A&A...292..641J} and \cite{1995A&A...300..503D}. The opacities are temperature dependent. The dust temperature is calculated with \texttt{RADMC-3D} using an initial thermal Monte-Carlo simulation, with the host star modelled as a blackbody and a point source. The star is also taken to be the only source of light within the system. We choose to run the simulation for $10^8$ individual photons. The blackbody wavelength spectrum is generated evenly in log space.

An example of an image generated for a protoplanetary disc with a 30 $M_\text{J}$ companion orbiting the host star at 10 au can be seen in the top left panel of Figure~\ref{fig:conv}. Importantly, an azimuthally asymmetric dark region can be seen extending out behind the companion, suggesting that the companion is directly casting a shadow onto the outer disc.

\subsection{Synthetic Observations}

\begin{figure}
    \centering
    \includegraphics[width= 1\columnwidth]{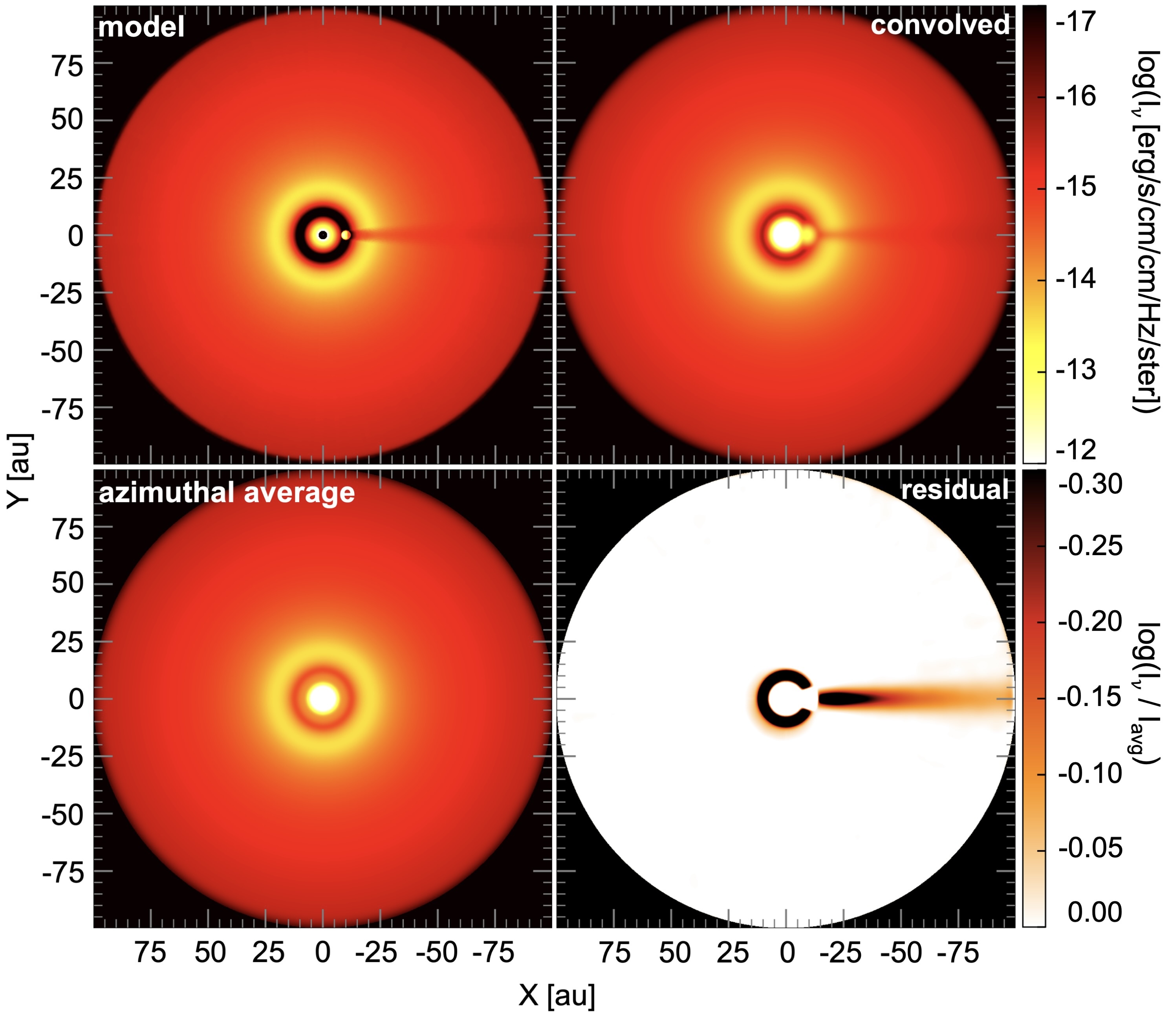}
    \caption{\textit{Top left:} Scattered light image (\texttt{RADMC-3D} output) of our disc model with a 30 $M_\text{J}$ companion orbiting at 10 au, simulated at $\lambda$ = 1 $\mu$m. \textit{Top right:} Convolution of \texttt{RADMC-3D} output with a PSF that has a FWHM of 0.04". \textit{Bottom left:} Azimuthally averaged version of the convolved image. \textit{Bottom right:} Residual image, obtained by dividing the convolved image by the azimuthally averaged image. The bottom colourbar is designated for the residual image only, while the other three images follow the top colourbar.
    }
    \label{fig:conv}
\end{figure}

\begin{figure}
    \centering
    {\includegraphics[width= 0.99\columnwidth]{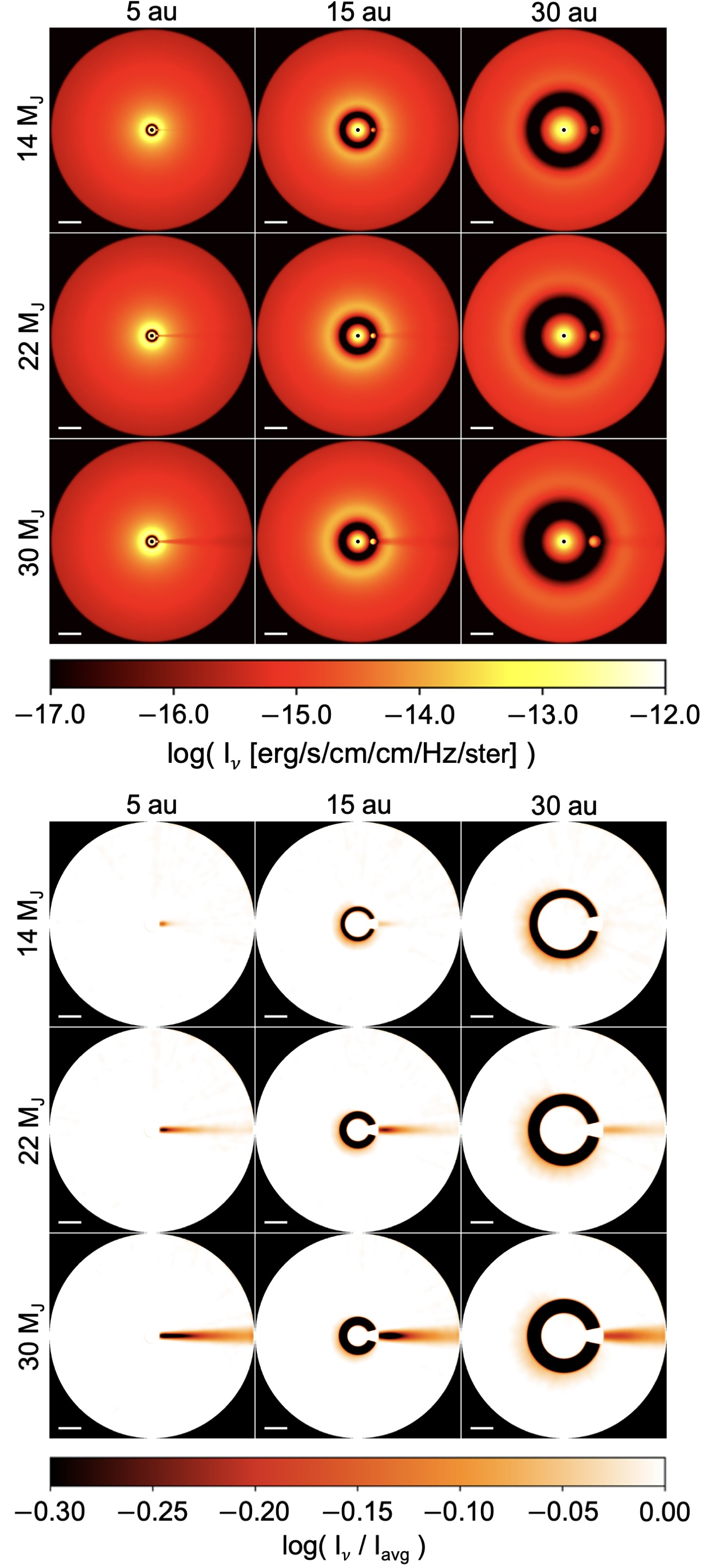}}
    \caption{\textit{Top:} Sample of synthetic intensity maps (scattered light images) produced by \texttt{RADMC-3D}. \textit{Bottom:} Corresponding maps normalised to the azimuthal average intensity. Columns indicate varying orbital distances while rows denote different companion masses. Each scale bar (bottom left of images) marks a distance of 20 au.}
    \label{fig:gallery}
\end{figure}

To use the model as a direct comparison to observed images we need to account for the instrument angular resolution of scattered light observations. SPHERE has an angular resolution of 0.035” \citep{2017Msngr.168...32K}; this finite resolution imposes a limit on the level of detail that can be obtained, particularly with respect to resolving substructures within the disc. We model a disc with a diameter of 200 au, which corresponds to an angular diameter of $\theta_\text{D} = 2$°, assuming a distance of 100 parsecs from the observer. To generate synthetic observations that incorporate the effects of finite angular resolution, we convolve our images with a point spread function (PSF) that has a full-width half-maximum (FWHM) of 0.04”. The top right panel of Figure \ref{fig:conv} provides an example of the convolution applied to a disc with a 30 $M_\text{J}$ companion located at 10 au. Following the convolution, some au-sized details are lost, particularly around the gap and Hill sphere. However, the shadow, although harder to discern visibly, remains detectable.
Appendix \ref{sec:convolve} discusses the difference between the convolved and unconvolved data in more detail.

\begin{figure*}
    \centering
    \makebox[\textwidth][c]{\includegraphics[width=1\textwidth]{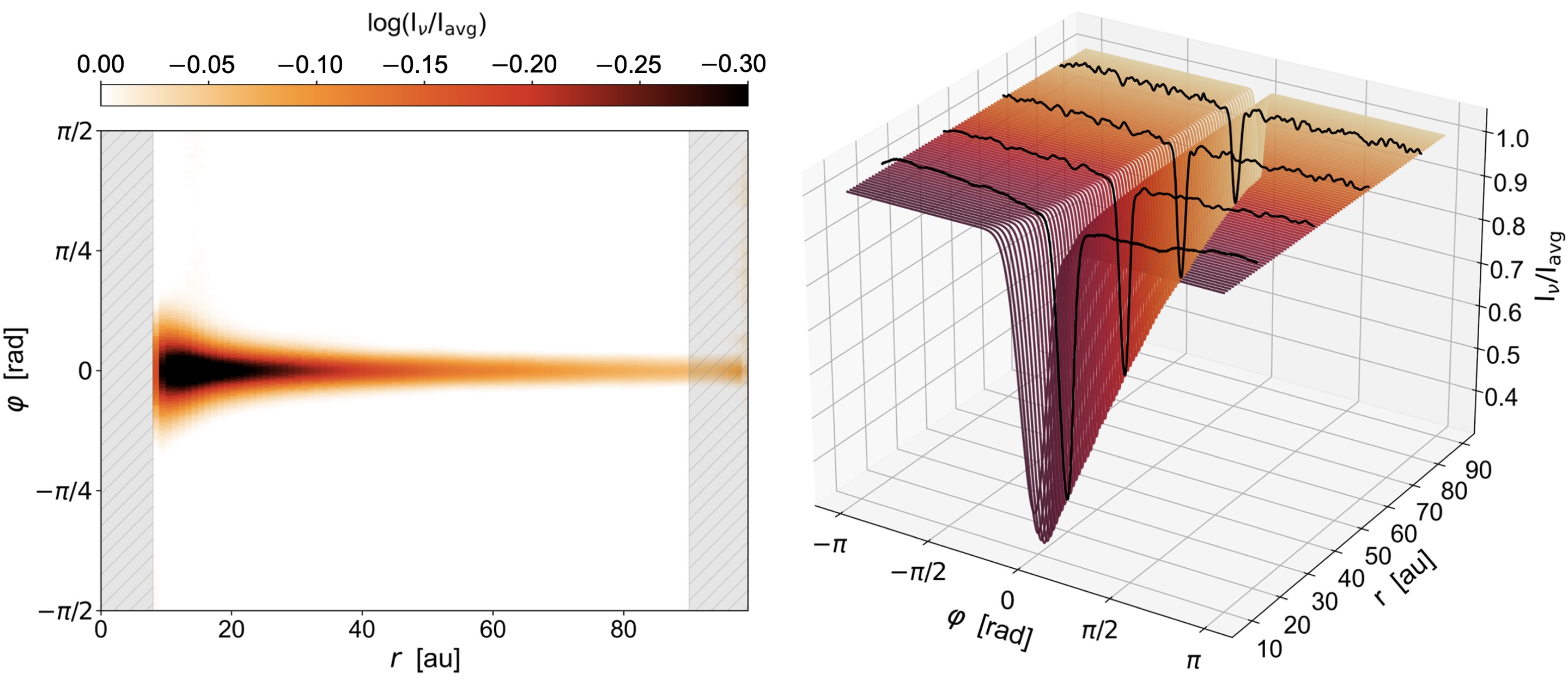}}
    \caption{\textit{Left:} Logarithmic intensity map of the deprojected disc, transformed from polar to Cartesian coordinates, where the x-axis is radius and the y-axis is the azimuthal angle, $\theta$. The map is normalised to the azimuthal average and displays the shadow caused by a 30 $M_\text{J}$ companion at 5 au. Shaded regions (0 < $r$ < 8 au and 90 < $r$ < 100 au) are excluded from the fit. Darker regions indicate greater negative deviations from the azimuthally averaged intensity at each radius. The shadow width corresponds to the angular extent along $\theta$. \textit{Right:} 3D Gaussian fit to the normalised intensity profile as a function of azimuthal angle and radius. The shadow depth appears as a dip along the z-axis, with colour denoting radial distance. The four black lines at $r$ = 20, 40, 60, and 80 au show examples of intensity data used for the Gaussian fits.}
    \label{fig:fitting}
\end{figure*}

\subsection{Azimuthal Averaging \& Normalisation}

Azimuthally asymmetric features within discs, such as shadows, can be highlighted by normalising the image against its azimuthally averaged counterpart. By dividing the convolved image (top right panel of Figure \ref{fig:conv}) by the averaged image (bottom left panel of Figure \ref{fig:conv}), we can isolate these asymmetric features, leaving behind a residual (bottom right panel of Figure \ref{fig:conv}). 

In the residual image, the black ring represents the gap, and the white cutout indicates the companion's location. The colour scheme denotes the relative deviation from the azimuthal average, with a pixel value of 0 meaning the region shows no asymmetric features. A lower pixel value indicates the region is darker than the azimuthal average at that specific annulus. As the focus is on the shadow, the displayed colour scale is limited to 0, meaning that regions brighter than the azimuthal average (e.g. the companion compared to the gap at the same annulus) are also displayed as white.

\section{Results \& Analysis} \label{sec:results}

A selection of synthetic scattered light images obtained from our simulations are shown in the top panel of Figure~\ref{fig:gallery}. The grid of disc surface brightness spans over the companion's location (columns) and mass (rows). The bottom panel, instead, displays the residual maps, after the post-processing (convolution and normalisation) has been applied. A full gallery of synthetic scattered light images and their corresponding residual maps is provided in Appendix~\ref{app:outputs}. 

We observe that more massive companions with smaller orbital separations cast a deeper shadow onto the outer disc, compared to smaller and more distant companions. Consequently, the most pronounced shadow within our parameter space is observed in the case of a 30~$M_\text{J}$ companion at 5 au. It should be noted that companions with masses below 14~$M_\text{J}$ were unable to produce a detectable shadow within the chosen disc geometry.

\begin{figure*}
    \centering
    \includegraphics[width=\textwidth,trim=0cm 0cm 0cm 0cm,clip]{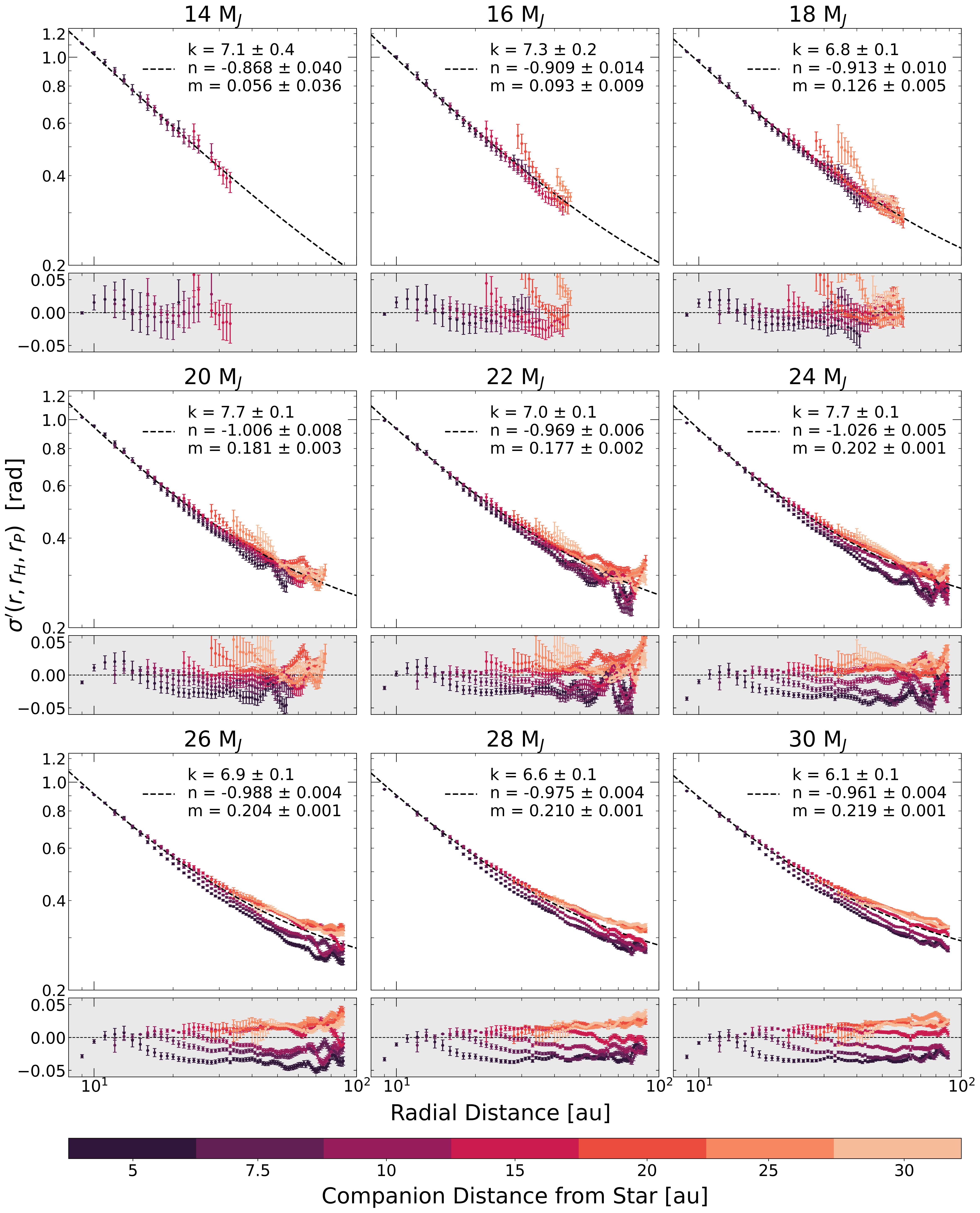}
    \caption{Scaled shadow width as a function of radius. Each panel shows the dataset for a specific companion mass, with colours indicating datasets for a fixed orbital separation. Error bars represent the uncertainty in the shadow
width from the Gaussian fitting. The dashed black line is the best-fit curve, whose parameters are provided in the legend. The grey shaded panels underneath show the residuals, calculated as the difference between the data points and the best-fit line.}
    \label{fig:width_grouped}
\end{figure*}

\subsection{Shadow Characterisation}

The decay of the shadow, both radially and azimuthally, changes with planet mass and distance, as seen in Figure ~\ref{fig:gallery}. The deprojected intensity map in the left panel of Figure \ref{fig:fitting}, where we display the normalised intensity for the case of a 30 $M_\text{J}$ companion at 5 au, shows that the shadow decreases in width and depth (decrease in relative intensity with respect to azimuthal average) radially. As in the lower panel of Figure ~\ref{fig:gallery}, the darker regions in the deprojected intensity map indicate greater shadow depth.

In order to quantify this behaviour, we fit a Gaussian function to the azimuthal intensity profile at each radius. The normalised intensity distribution can be modelled as, 
\begin{equation}
    I_{\nu}(r, \theta) = 1 - A(r) \; \rm{exp}\left[ - \frac{1}{2} \left( \frac{\theta}{\sigma(r)} \right)^{2}  \right], 
\end{equation}
where $A(r)$ is the amplitude of the Gaussian and is taken as the depth of the shadow. The standard deviation, $\sigma(r)$, corresponds to half of the shadow width. Since the intensity is normalised, we assume that the azimuthally averaged (background) intensity is constant at a value of 1. 

We establish specific fitting criteria for the data; first, we require that the decrease in intensity be at least 5\% ($A(r) \geq 0.05$) in order to confidently distinguish it from noise. Additionally, we impose an empirically motivated constraint of $\sigma \leq 2$ rad, based on our observation that among all of our simulations, the intensity dips identified as shadows were well below this width limit, even in the case of large companions close to the star, hence subtending a large angle. This constraint is particularly relevant when fitting intensity profiles at large radii, where the profile may appear nearly flat, leading to erroneous Gaussian fits over the whole azimuthal range. 

This constraint on width is also relevant at very small radii, right after the companion location. We find that at these distances, the intensity shows varying behaviour that cannot be well described by a Gaussian function. This may lead to the least-squares algorithm to fit data that doesn’t exhibit a well-defined Gaussian shape, which we prevent with the previously mentioned constraint on the width. Instead, we begin the Gaussian fitting at the radial distance where the dip in intensity first becomes well described by a Gaussian function. This distance is empirically found for each sample and hence varies depending on companion mass and orbital distance. We choose to have a hard cut-off for the fitting at 90 au, since the normalised intensity beyond this point is not well represented by a Gaussian due to the geometry of the disc model. This could be due to the weakening of the signal, making it difficult to achieve an accurate fit. Furthermore, back-scattering of photons from the 100 au boundary begin to introduce unphysical artifacts in the shadow profile.

For the case of a shadow cast by a 30 $M_\text{J}$ companion orbiting at 5 au, we display the 3D plot of the fitted intensity profile, as a function of radius and azimuthal angle, in the right panel of Figure \ref{fig:fitting}. Following the fitting criteria listed above, the fitting begins at 8 au in this instance, revealing a steep intensity drop at the companion's azimuthal position ($\varphi$ = 0). The shadow depth can be visualised as the dip in the z-axis and is seen to decrease moving radially outwards. The shadow width is greatest near the companion and narrows radially outward. The colour scheme denotes radial distance, with the tone getting lighter with increasing radius. In addition to the fitted Gaussians, we plot the normalised intensity data at the radial distances of $r$ = 20, 40, 60, and 80 au, in black. As it can be seen, the Gaussian distributions accurately fit the data, with the main deviations between data and model being observed at the flat tails of the Gaussians. This is also supported by reduced Chi-squared values for the fits; we find the average reduced Chi-squared value for this dataset to be $\langle \chi^2_{\nu} \rangle =1.26$, suggesting a reasonable fit across the models.

\subsection{Shadow Width} \label{sec:width}

As seen in Figure~\ref{fig:fitting}, the width of the shadow decreases with distance from the star. Additionally, we find that the shadow's features depend on the companion's properties, including its mass and orbital distance. Therefore, we choose to scale the shadow half width by the companion's Hill radius using the relation,
\begin{equation}
    \sigma'(r, r_\text{H}, r_\text{p}) = \sigma(r) \cdot \left( \frac{r_\text{H}}{r_\text{p}} \right)^{-1},
\label{scaling}
\end{equation}
where the scaling term is a proxy for the angle subtended by the Hill sphere at the planet's orbital distance, $\theta = 2r_\text{H}/r_\text{p}$. This term also encompasses the dependence on the companion's mass, $\propto (M_\text{p})^{1/3}$.

Initially, we group the data with respect to companion mass, and present the results in Figure~\ref{fig:width_grouped}. Each panel displays data for shadows caused by a specific companion mass. The colours within each plot indicate the different companion orbital distances. The error bars represent the uncertainty in the shadow width arising from the Gaussian fitting in Section ~\ref{sec:width}. We find that each dataset is well-described by a power law of the form, 
\begin{equation}
    \sigma'(r, r_\text{H}, r_\text{p}) = k \cdot r^{n} + m,
\label{eq: width}
\end{equation}
where the values of $k$, $n$ and $m$ are determined via least-squares fitting. The curve of best fit is plotted in a black dashed line, and the best fit parameters are listed in the top right corner, with corresponding fitting uncertainties. While parameters $k$ and $m$ vary between each dataset, we find that the exponent, $n$, stays consistent around -1. This means the shadow width scales as an approximate inverse power law of the form $\sigma(r)' \propto 1/r$. We see from the data that the shadow width declines rapidly at the onset but this decline slows down as we move further out in the disc; this matches what we see in Figure \ref{fig:fitting} where the shadow width initially decreases quickly, then appears to plateau at larger radii.

Below each panel, we plot the corresponding residual plots - calculated as the difference between the data points and the best fit curve. We see that the variability within each dataset is bound to $\pm 0.1$ of the curve, indicating that the power law is a good fit across all masses. We see, especially for the lower mass companions ($14 \geq M_\text{J} \geq 24$), that the data cannot be conclusively be characterised for different companion distances within each mass bin, even though some layering is observed for higher masses ($26 \geq M_\text{J} \geq 30$).

\begin{figure}
    \centering
    \includegraphics[width=\columnwidth, trim=0cm 0cm 0cm 0cm,clip]{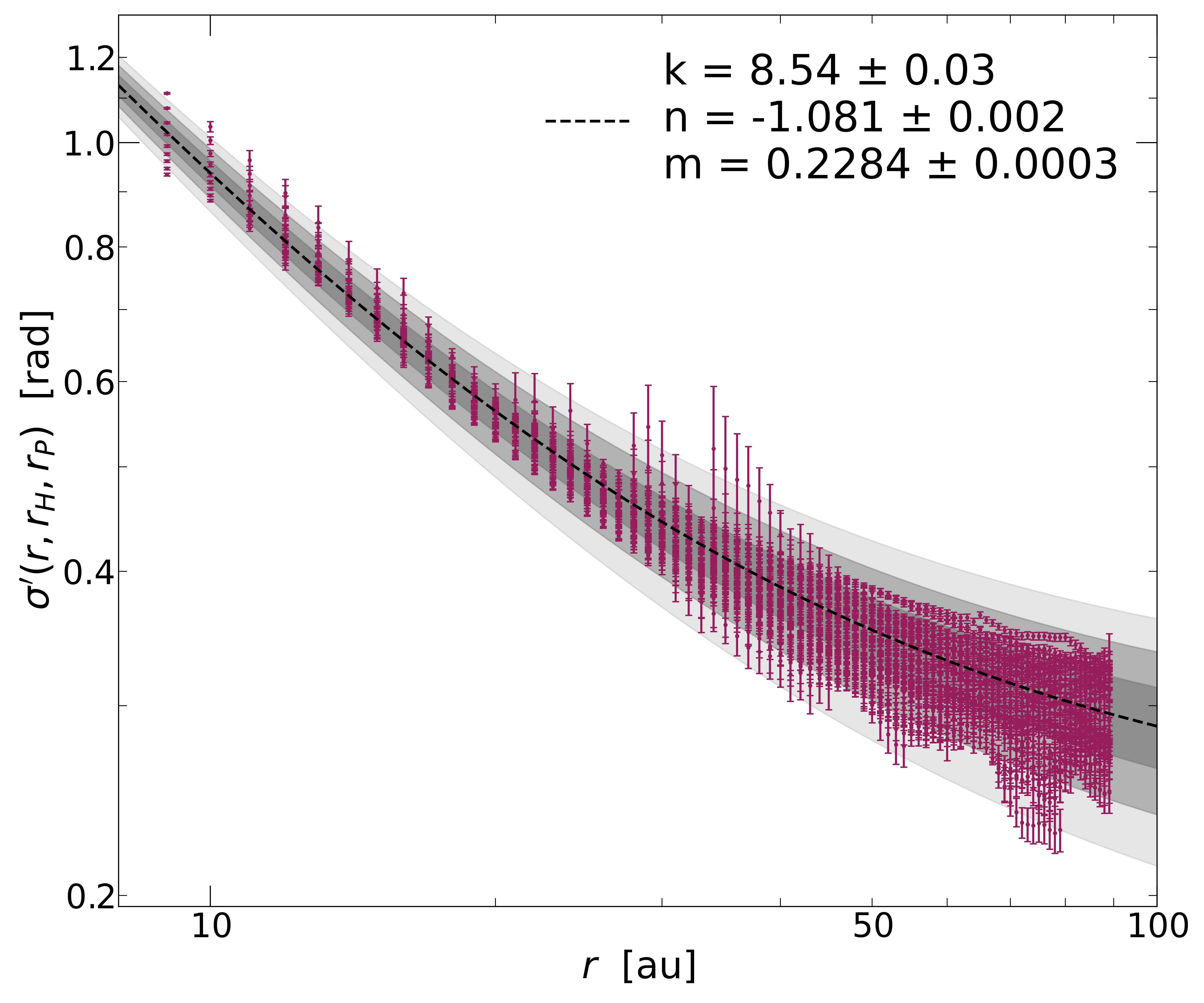}
    \caption{Scaled shadow width as a function of radius across the total dataset (log-log scale). Error bars show uncertainties from Gaussian fitting. The dashed black line represents the best-fit curve, with parameters given in the legend. Grey shaded bands indicate regions within $\pm 1\sigma$, $\pm 2\sigma$, and $\pm 3\sigma$ of the data, with lighter shades denoting higher sigma levels.}
    \label{fig:width_total}
\end{figure}

Since each mass-separated dataset shows an approximate inverse power law behavior, we plot the scaled shadow width as a function of radius for the combined dataset in Figure ~\ref{fig:width_total}. Notably, the trend across the entire dataset also follows $\sigma(r)' \propto 1/r$, indicating that the decrease in shadow width with radial distance can be modeled by an inverse power law, independent of companion mass or location.

To quantify data spread, we calculate the standard deviation of the residuals, defined as the difference between each data point and the best-fit curve. This standard deviation is found to be $\sigma = 0.025$ (3 s.f.). Approximately 74.8\% of all fitted widths fall within $\pm 1\sigma$ of the curve, 98.1\% within $\pm 2\sigma$, and 99.86\% within $\pm 3\sigma$, indicating a high degree of conformity between the width distribution and the fitted model. These tolerance bands are illustrated with grey shading in Figure~\ref{fig:width_total}, with darker shading indicating lower $\sigma$ values. Notably, the \textit{relative} deviation—as a percentage of the curve value—becomes more pronounced at larger radii, where shadow width values are smaller.

\begin{figure*}
    \centering
    \includegraphics[width=\textwidth,trim=1.5cm 49cm 1cm 1cm,clip]{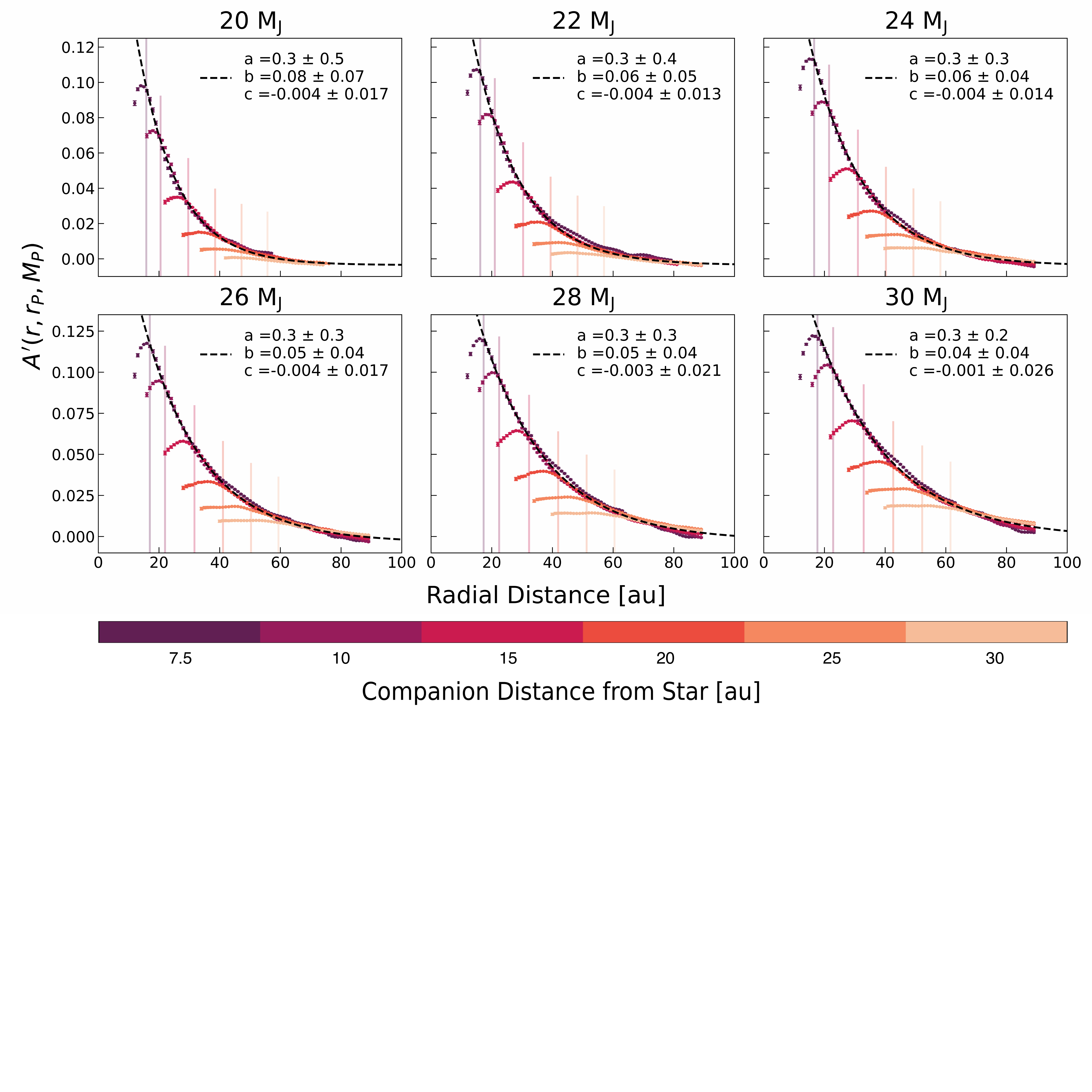}
    \caption{Scaled shadow depth, $A'(r, r_\text{p}, M_\text{p}$), as a function of radius. Each individual panel displays the dataset for a specific companion mass, with colours indicating different companion orbital separations. Error bars represent the uncertainty in the amplitude calculated from the Gaussian fitting. The same scaling is applied for all panels, according to Equation~\ref{eq:amplitude}. The solid vertical lines mark the gap outer edge for each companion. The dashed black line is the best-fit curve to the decaying tails of each distribution. The parameters describing the curve, following Equation~\ref{ref:exponential}, are displayed in the top right corner of each panel.}
    \label{fig:amp}
\end{figure*}

\subsection{Shadow Intensity}
The shadow intensity can be analyzed by observing how the amplitudes of the fitted Gaussians change with increasing radial distance. Similar to the approach taken for shadow width, we group our data based on companion mass and use different colors to represent data from various orbital distances. The results are shown in Figure \ref{fig:amp}. In contrast to the shadow width, the variation in shadow intensity follows a more complex pattern: the shadow depth initially increases, reaches a local maximum, and then decreases. This is also evident in the left panel of Figure \ref{fig:fitting}, where the darkest point does not occur at the very inner edge of the shadow but at a slightly larger distance.

We observe that the outer edge of the gap (i.e., $r = r_\text{p}+\Delta_{\text{gap}}/2$) approximately corresponds to the threshold beyond which the intensity monotonically decays. These points are indicated by the vertical lines in each panel. It is unclear whether the initial Gaussian-like peak is a result of physical phenomena or if it should be fully attributed to the convolution process in which we convolve the data with a PSF.

To compare the datasets with the same companion's mass, we scale the amplitudes so that the decaying tails align with one another. This alignment is achieved in three steps. Firstly, we limit the data to only the monotonically decreasing region, meaning those points beyond the gap's outer edge. We then analyse each mass grouping independently. We understand that differences among each dataset are likely influenced mainly by the orbital distance of the companion. Consequently, we decide to scale each dataset of a given mass as 
\begin{equation}
    A'(r, r_\text{p}) = \left[ A(r) - \alpha \left( \frac{r_\text{p}}{\textrm{1 au}}  \right)^{\beta} \right] \times \left( \frac{r_\text{p}}{\textrm{1 au}}\right)^{\gamma}.
    \label{eq:scale}
\end{equation}
The optimisation process iterates through different combinations of the scaling parameters $\alpha$, $\beta$, and $\gamma$ and seeks to minimise the sum of squared differences between the scaled datasets. This is repeated for each radial point, and the final scaling parameters are selected based on the overall reduction in variance within the mass grouping.

At this stage, the data within each mass grouping is aligned, however the scaling parameters are different for each grouping. In order to find a common scale for all mass groupings, we look at how $\alpha$, $\beta$, and $\gamma$ individually evolve as functions of mass and determine empirical relations. We find via least-squares fitting that $\alpha \propto M_\text{p}^{10}$, $\beta \propto M_\text{p}$ and $\gamma$ is a constant. The final empirical common scale for the amplitude of the shadow is hence found to be:
\begin{equation}
    \begin{split}
    A'(r, r_\text{p}, M_\text{p}) = & \left[ A(r) - 8.3 \times 10^{-16} \left( \frac{M_\text{p}}{M_\text{J}} \right)^{10} \left( \frac{r_\text{p}}{\textrm{1 au}}  \right)^{-0.1 \frac{M_\text{p}}{M_\text{J}} + 2.5} \right] \\ 
    & \times \left( \frac{r_\text{p}}{\textrm{1 au}}\right)^{-0.8}.
    \end{split}
    \label{eq:amplitude}
\end{equation}
This scaling is applied to all mass groupings, with the results shown in Figure \ref{fig:amp}. It should be noted, that for clarity we choose to plot the entire datasets, which show that the decaying tails do indeed overlap, while the initial peaks do not.

\begin{figure*}
    \centering
    \includegraphics[width=1\textwidth,trim=0cm 0cm 0cm 0cm,clip]{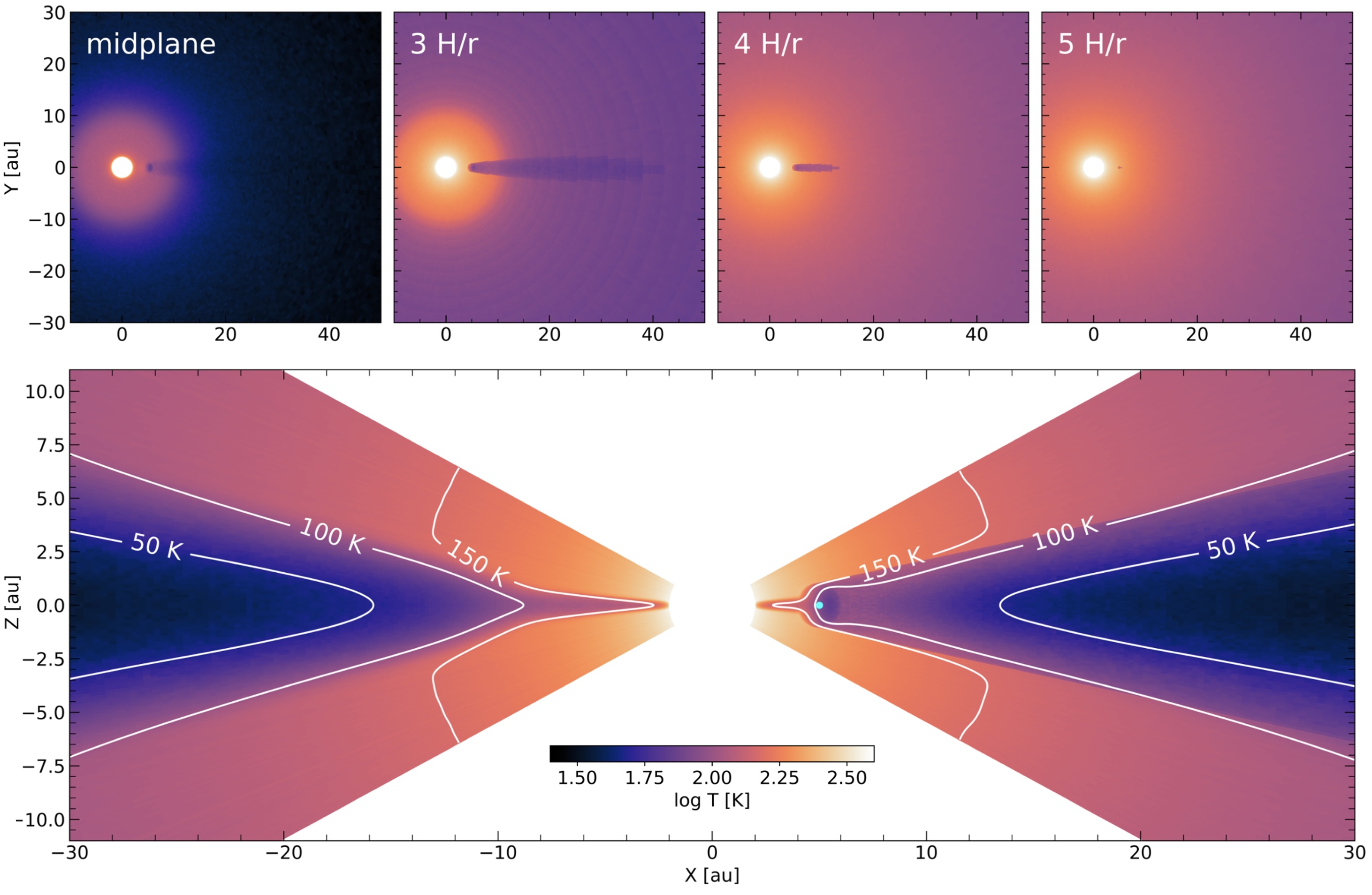}
    \caption{Temperature maps of a disc with a 30 $M_\text{J}$ companion at 5 au. \textit{Top:} Face-on view of the dust temperature distribution, illustrated at the midplane, 3H/r, 4H/r and 5H/r in each panel, respectively. \textit{Bottom:} Cross-section of the temperature map along a cut at the azimuthal position of the planet. The right side of the disc shows the cross-section at the companion azimuth ($\varphi_{P}$), while the left side is at the opposite azimuth ($\varphi_{P} + \pi$). The white contours highlight the isothermal surface and the cyan dot indicates the companion location.}
    \label{fig:temperature1}
\end{figure*}

\begin{table}
    \centering
    \caption{Total averaged fitting coefficients for Eq. \ref{ref:exponential}. The values are presented with their associated uncertainties.}
    \begin{tabularx}{\columnwidth}{c>{\hspace{1.3cm}}c>{\hspace{1cm}}c}
        \toprule
        $\langle a \rangle$ & $\langle b \rangle$ & $\langle c \rangle$\\
        \midrule
        $0.29 \pm 0.14$ & $0.055 \pm 0.019$ & $-0.0034 \pm 0.0076$ \\ 
        \bottomrule
    \end{tabularx}
    \label{tab:depth}
\end{table}

We observe that the decaying tails of each distribution follow an exponential decay. After applying the scaling process, we fit the "normalised" amplitudes using the functional form,
\begin{equation}
    A(r)' =  a \cdot e^{- b \cdot r} + c,
    \label{ref:exponential}
\end{equation}
where r represents the radial distance, and a, b, and c are the fitting parameters. This fitting procedure is performed individually for each mass-grouped dataset. The resulting best-fit parameters are displayed in the top-right corner of each plot. We find that the fitting parameters for different mass groupings overlap within their respective uncertainties. As such, we list the total averaged coefficients for the shadow depth fitting in Table~\ref{tab:depth}. Using these averaged coefficients along with Equation \ref{eq:amplitude} and Equation \ref{ref:exponential}, the depth of a shadow in a disc can be predicted based on companion mass and orbital distance. While the coefficients a and c appear largely independent of mass, the coefficient b may show a weak dependence, though it remains constrained within the range 0.04–0.08, and it is consistent within the margin of uncertainty for all models. 

We underline that this scaling process is not applied to the data from the lower mass range \( 20~\textrm{M}_J \leq \textrm{M}_\text{p} \leq 30~\textrm{M}_J \), as there is insufficient overlapping data within each dataset of a given mass to effectively determine the scaling parameters. Furthermore, we note that the specific case of shadows cast by a companion orbiting at 5 au notably does not fit the pattern as well as the rest, and hence is not taken into account when scaling the data. 
The convolution step smears out the most narrow shadow lanes, such that they appear shallower. Therefore, for companions inside 10 au, the shadows are not resolved and do not follow the fit as the resolved shadow features do. Qualitatively this caveat also applies to observational data in the same way.

\subsection{Temperature in the Disc}

\begin{figure}
    \centering
    \includegraphics[width=\columnwidth]{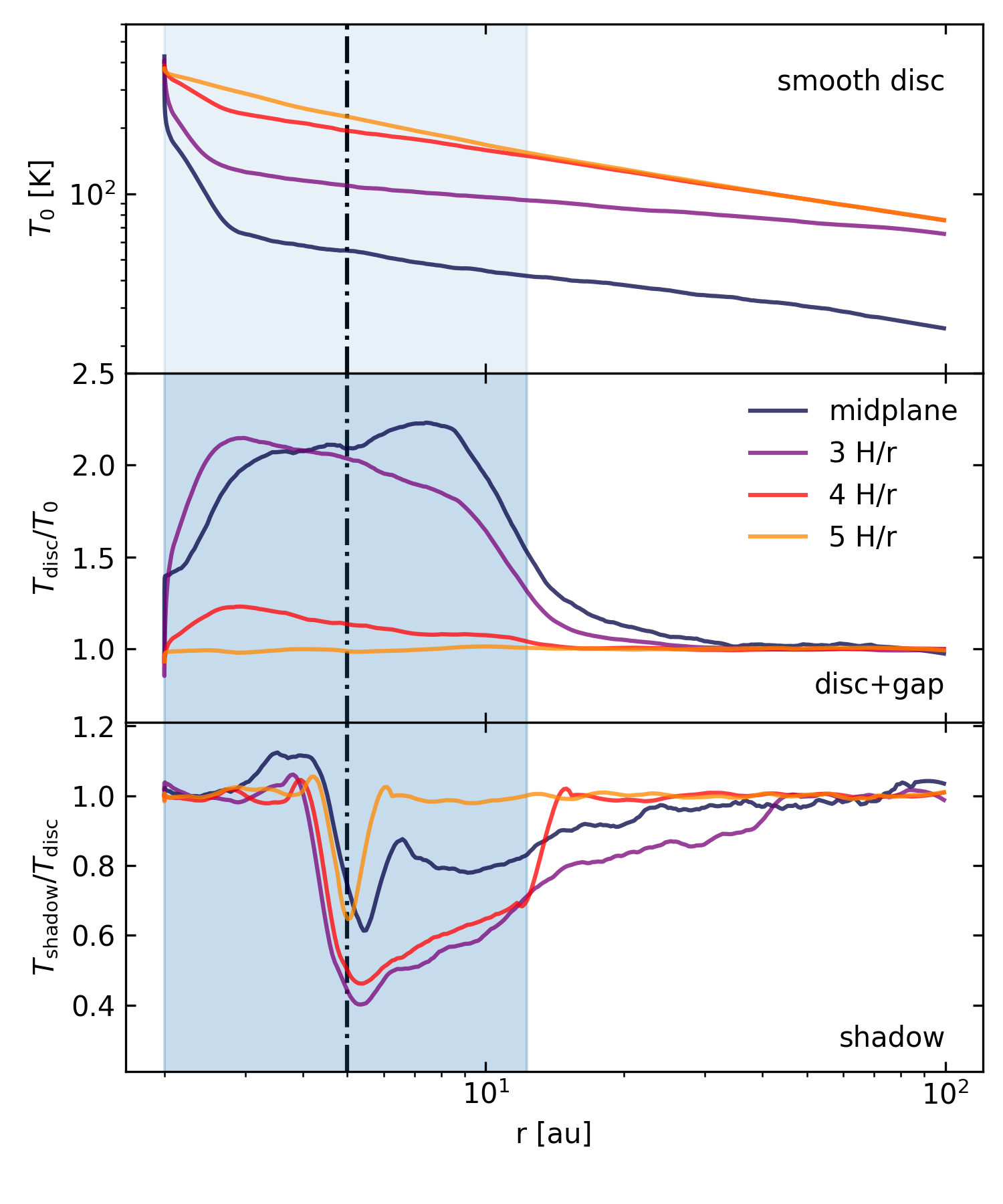}
    \caption{Radial dust temperature profiles. \textit{Top}: Temperature profile of a smooth flared disc without a companion or a gap feature, $T_0$. \textit{Middle}: Temperature profile of the disc with a gap, $T_\mathrm{disc}$, normalised to the smooth disc temperature profile. \textit{Bottom}: Temperature ratio between the gapped disc, $T_\mathrm{disc}$, and the  temperature profile at the azimuthal position of the companion, $T_\mathrm{shadow}$. The results are shown for a 30~$M_\text{J}$ companion located at 5~au (indicated by the black dashed line). The different colour lines represent the different disc scale heights, which are the same as illustrated in Figure~\ref{fig:temperature1}. The shaded area in each plot represents the gap width.}
    \label{fig:temperature2}
\end{figure}

Current observations of protoplanetary discs have detected azimuthal variations in both the gas and dust surface density, implying differences in the temperature distributions \citep[e.g.,][]{2023A&A...670A.154W, 2023NatAs...7..684K}. Motivated by this, we look at the impact of the companion on the disc's temperature structure. It is essential to recognise that our model is static; it assumes the companion remains stationary, does not add a thermal contribution and allows the disc to fully adjust to this condition. 
This means our models produce the maximal temperature drop in the shadowed region, corresponding to an instantaneous cooling of the disc. This also neglects the thermal interaction of the companion with its environment through irradiation or shock heating.
Figures~\ref{fig:temperature1} and \ref{fig:temperature2} show that the dust temperature distribution within the disc varies with height. Below the optical surface where the opacity exceeds 1 ($\lesssim 4 H/r$) \citep{2001Dullemond} the dust opacity shields the disc from direct irradiation and so the temperature decreases towards the midplane to values similar to the temperature reported in observations \citep[e.g.,][]{2021MAPS}. 

What appears to have a prominent impact on the disc temperature structure is the presence of the gap carved out by the companion. Since the gap is an optically thin, directly irradiated region, temperatures within it are higher than temperatures in a gap-free, smooth disc, $T_0$, (top panel of Figure \ref{fig:temperature2}). The middle panel of Figure \ref{fig:temperature2} shows that this effect is particularly evident in the midplane where temperatures reach up to 2.25 times higher than those in a smooth disc. This elevated temperature within the gap may have important implications for the formation and distribution of materials in the disc.

\begin{figure}
    \centering
    \includegraphics[width=\columnwidth]{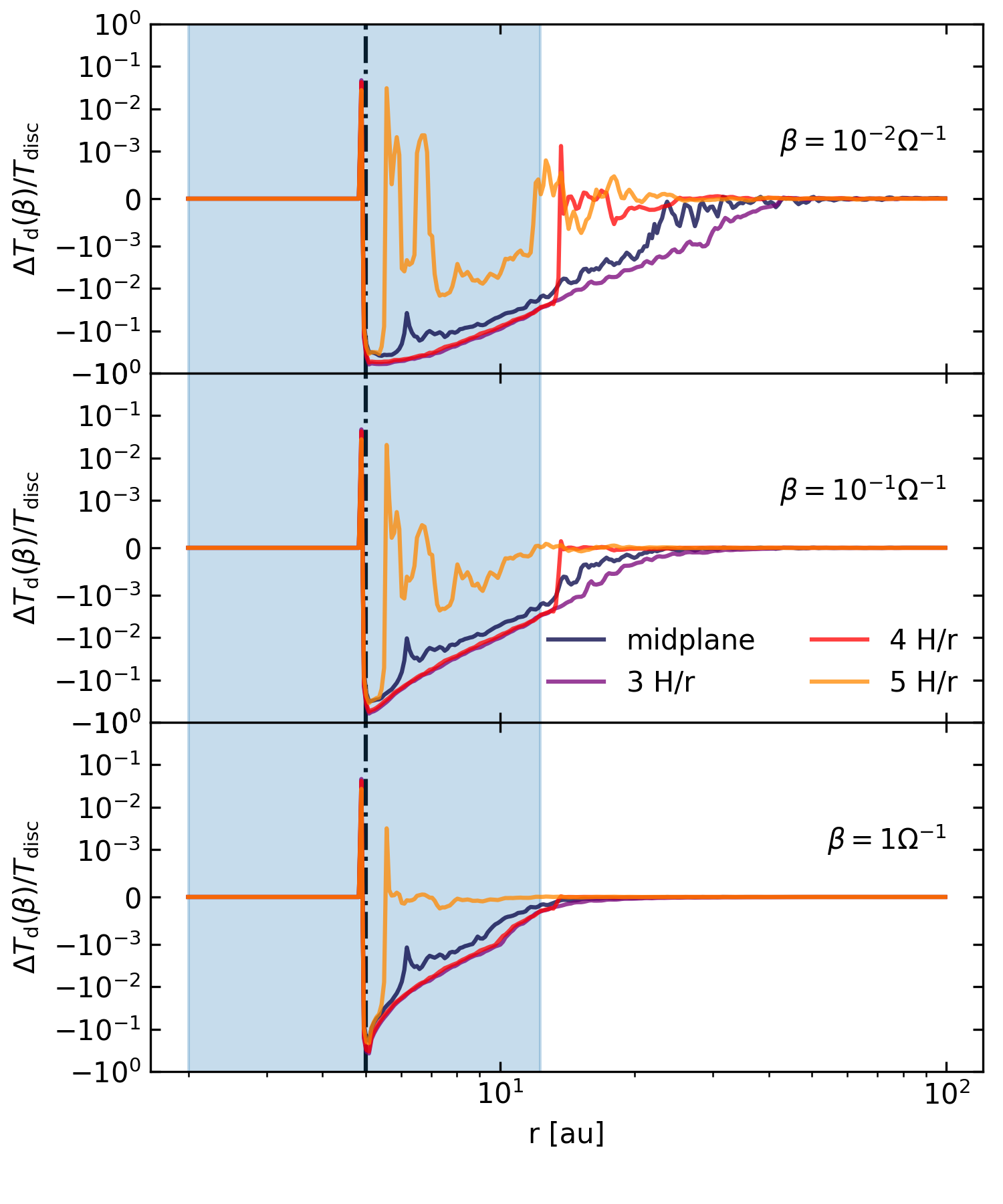}
    \caption{Temperature reduction $\Delta T_\text{d}$ relative to the unshaded disc temperature $T_\text{disc}$ as a function of cooling time $\beta$, shown at different vertical heights within the disc. Cooling times decrease from bottom to top: 1, 0.1, and 0.01 $\Omega_\text{K}^{-1}$, where $\Omega_\text{K}^{-1}$ is the unit of the radially dependent Keplerian orbital period. Line colours correspond to different vertical heights, measured in units of the local scale height of the disc, $H/r$.}
    \label{fig:beta_cool}
\end{figure}

The bottom panel of Figure~\ref{fig:temperature1} shows a cross-section of the disc temperature map, with the companion location marked by a cyan dot. Comparing the right side (along companion azimuth) and the left (opposite to companion), we see that the presence of the companion reduces the temperatures of the dust radially
within the disc. For instance, with only the gap profile present, the midplane cools to 100 K at approximately 10 au. However, along the companion's azimuth, this cooling occurs at about 5 au, effectively halving the distance needed to reach 100 K. This demonstrates that the dust surrounding the companion effectively blocks direct irradiation from heating the disc behind the companion, resulting in a "thermal shadow" where temperatures are lower than they would be in the absence of the companion. Figure \ref{fig:temperature1} shows the extent of this shadow varies with height within the disc. We observe the most substantial temperature drop at $3~H(r)/r$, where the tail of the cooled region extends to 40 au. This is also visible in the bottom panel of Figure \ref{fig:temperature2}, which shows the radial temperature profile along the companion’s azimuth, $T_{\text{shadow}}$, normalised to the radial temperature along the opposite azimuth, $T_{\text{disc}}$. At $3~H(r)/r$, the companion’s static presence can reduce temperatures by up to 30\% relative to $T_{\text{disc}}$ within the disc. Above the optically thin limit, the companion’s impact on disc temperature diminishes, both in terms of the cooled region’s spatial extent and the degree of temperature reduction. At $4~H(r)/r$ the cooled region only extends to $\sim$15 au and by $5H(r)/r$ the temperature dip is confined within the Hill sphere, with $T_\text{shadow}$ equal to $T_{\text{disc}}$ by $\sim$6 au.

\subsubsection{Dynamic Temperature Reduction}

Considering a dynamic system where the companion moves along its orbit, the temperature reduction the disc may experience when shaded can be estimated based on the relative shadow coverage over one orbit, 
\begin{equation}
t_\mathrm{dark} = \frac{2 \sigma(r)}{2 \pi r} \frac{1}{\Omega_\mathrm{K}(r) - \Omega_\mathrm{K}(r_\text{p})}, 
\end{equation}
and the cooling time required for the system to adjust to a new thermal state, $\beta$, which is measured in units of the orbital timescale $\Omega_\mathrm{K}^{-1}$. The factor of $\Omega_\mathrm{K}(r) - \Omega_\mathrm{K}(r_\text{p})$ represents the relative orbital motion between the disc and the planet. 
The temperature reduction with time $ \partial T_\text{d}/ \partial t$ can be approximated by,
\begin{equation}
\frac{\partial T_\text{d} (r)}{\partial t} \approx \frac{\Delta T_\text{s}(r)}{\beta},
\end{equation}
where $\Delta T_\text{s} (r)$ is the static (upper limit) in temperature reduction, $\Delta T_\text{s}(r) = T_\text{disc} (r)-T_\text{shadow}(r) $. Critically, this estimate ignores the Gaussian depth variation of the shadow and corresponding azimuthal temperature variations as well as any heat transport between the heights during cooling. 
We now combine the time the gas spends in the shadow, $t_\text{dark}$, with the decay timescale, $\beta$, to determine the temperature reduction within the shadow, $\Delta T_\text{d}$, as follows
\begin{equation}
\Delta T_\text{d} = - \Delta T_\text{s}(r)(1 - \exp{(- t_\text{dark}/\beta )}).
\end{equation}

In Fig.~\ref{fig:beta_cool}, we estimated the amount of cooling that would occur for various cooling times.
The cooling time in the outer disc depends on the local opacities and dust properties, but it can reach a value of $\beta \approx 0.1 - 1 \Omega_\text{K}(r)^{-1}$. 
Using a shadow coverage of $5-10\%$ and a static temperature reduction between $20-30\%$, we find that $\Delta T_\text{d} (r)$ is of the order of $\sim 0.02 - 0.5\%$ at a height of $3 H/r$. The amount of cooling will depend on the radial distance.
A dynamic model would provide a more comprehensive understanding of how temperature variations evolve over time and could determine whether the cooling effect of the shadow  persists long enough to cause lasting impacts on disc processes. 
Furthermore, dynamic-radiative feedback from the shadow can induce variations in the optical surface height of the disc, potentially leading to temperature increases in its upper layers \citep{2023Muley}. These temperature fluctuations can significantly affect CO observations, whereas, in scattered light, the shadow itself remains the most prominent feature.

However, this level of analysis lies beyond the scope of our static model.

\section{Discussion} \label{sec:discussion}

While the simplicity of our model allows for flexibility and consistent analysis, it inevitably introduces certain assumptions and approximations. The first key assumption involves the single gap created by the embedded companion. Observations of protoplanetary discs often reveal narrow gas rings, indicating low-viscosity environments \citep{2018Dsharp6, 2022Villenave}, which tend to favour deep, circular gaps \citep{2006Crida}. However, the profile of a gap can vary depending on the disc's thermodynamic conditions, and planets may even form multiple gaps \citep[e.g.][]{2020Ziampras}. Additionally, the companion masses in our model exceed the planetary mass regime on which the \cite{kanagawa_2016} gap model is based. In higher mass regimes, hydrodynamic simulations show temporally varying \citep{2022Scardoni} and eccentric \citep{DurmannKley2015, 2021Dempsey} disc profiles, as the companion-to-star mass ratio approaches that of a binary system. Although our static model may not capture these dynamical features, we opt for this approach to isolate the shadowing effect without interference from additional structures. We believe the signal in the disc immediately beyond the companion is significant enough to test observed intensity reductions against our model.

In our model we use the Hill sphere of the intended planet as the shadow causing Gaussian density distribution. This choice is a simplification from the complex mechanism of the 3D process of planetary accretion. \cite{2024Lega} has shown accretion of material from above the planet which can be optically thick at 0.2 Hill sphere above and around the planet.
While our model does not include a complex description of the accretion structure, the most important component is the size of the bound sphere of material. Our assumption of the complete Hill sphere produces the maximal effect, however, in our model mass and size of the material sphere are interchangeable with $r_\mathrm{H}\propto M_\mathrm{p}^{1/3}$ and the model can be rescaled to assume a different mass-to-size ratio.

Our model assumes that dust particles remain perfectly coupled to the gas and are suspended at multiple pressure scale heights, allowing us to use Equation \ref{eqn: vertical distribution} to describe the dust density distribution. However, in reality, dust settling \citep{2009Fromang} can lead to a sharper taper at around 3–4 pressure scale heights, even for small grains. A full treatment of settling is further complicated by the influence of radiation pressure, which can shape the vertical structure of the disc in non-trivial ways \citep[e.g.][]{2024Robinson}. By neglecting these effects, our model represents a conservative scenario where the disc remains more vertically extended than it might in a more detailed calculation. Importantly, this neglected settling reduces the amount of dust at high altitudes, effectively lowering the altitude of the optical surface. As a result, shadows cast by a companion could be more pronounced and extend farther than our current predictions suggest, making them easier to detect.

We use amorphous olivine as dust species, however as we look at large radii, ices can become important as well. The temperatures within the disc change the grain structure, composition and sizes throughout the disc, which affects the opacity and scattering of the grains \citep[e.g.][]{2022Tazaki}. Such effect would be mostly radially dependent and are mostly corrected for by considering the azimuthal average in the reduction, but could lead to a varying pattern in the deepest shadow region in the residual.

Our study gives theoretical upper limits on the size and depth of companion shadow, which can be used to reevalute reduced light features in disc with known companions such as PDS 70 \citep{2024PDS70} or HD 100453 \citep{2017Benisty} and investigate if other features can be consistent with a companion like in HD 169142 \citep{2020Bertrang, 2022Poblete}.

Through our suite of radiative transfer simulations, we find that the shadow narrows radially behind the companion following a $1/r$ relationship, rather than a fixed solid angle. This behaviour arises from a combination of factors, including the shape of the obscuring material, the disc’s flaring slope, and scattering effects. While a first-order analytic prediction for this result may be possible, deriving such a relation is beyond the scope of this study. Instead, we find it instructive to compare our radiative transfer treatment with a zeroth-order geometric approximation of the problem. 
Appendix \ref{sec: geometric} provides further details on the differences between geometric and radiative transfer predictions, including how scattering alters shadow morphology across various disc flaring indices. We show that geometric models notably alter shadow width at small radii, where scattering effects become more significant.

With further work, it could be shown that a shadow exhibiting a $1/r$ radial dependence may serve as a strong indicator for the presence of a companion, as opposed to other shadow causing phenomena. Our current analysis shows that the full dataset aligns within $\pm 3 \sigma$ of the best-fit curve. Notably, the width scaling factor we used depends on mass ($r_\text{H}/r_\text{p} = \sqrt[3]{M_\text{p}/3M_{*}}$), this could potentially serve as a new method for inferring the mass of a companion based on shadow observations that fit the $1/r$ prescription.

Another consideration is whether datasets grouped by companion mass can be used to determine how companion properties influence the $1/r$ decay. We find that when grouped by mass, the data exhibits a tight fit to the model, with a spread of $\pm 0.05$ from the curve. Although some ``layering'' can be observed, particularly in the higher-mass residuals in Figure \ref{fig:width_grouped}, it’s unclear if this deviation, based on orbital distance, warrants introducing an additional degree of freedom. One could speculate that the layering might result in a possible parameterisation of the width equation with orbital distance. However, given the close alignment of data within each grouping, the effect is not pronounced enough to support a strong case for this.

Understanding the temperature structure within protoplanetary discs is important as it affects various physical and chemical processes. For example, complex organic molecules (COMs) typically form on the surfaces of icy grains in the cooler regions of the disc \citep{Walsh_2014}. If these grains migrate into the hotter environment of the gap, the COMs and other ices within their mantles can readily evaporate into the gas phase, increasing their abundances. This gas-phase material may then be accreted onto an orbiting companion within the gap. Additionally, the gap presents a unique thermal structure, with temperatures ranging from the water ice line (150 K) to the tar line (350 K – 530 K). According to \cite{Lodders_2004, 2022Bitsch, Mousis_2024}, any carbonaceous material formed through nonequilibrium processes or originating from the ISM would remain solid in this region of the disc. This allows such material to drift towards the gap efficiently, accumulate there and be incorporated into the companion, potentially providing a pathway for carbon-rich material to enrich forming bodies in the gap.

Other processes such as gas-phase chemistry, thermal desorption, photodesorption and X-ray desorption could also be affected by variations in temperature \citep{Walsh_2010}. Even small variations can alter the dynamics in the disc and lead to additional disc features as \cite{2024Zhang} and \cite{2024Ziampras} demonstrates. Therefore, the effect of the companion is not only crucial for interpreting disc observations but also has implications for the disc structure, warranting further investigation in future studies.

In our model, we assume the companion and its surrounding material are only heated passively, neglecting any accretion heating of the companion and shock-heating from the companion planet interaction. This simplification underestimates the companion's temperature, which could rise to $\sim 1000$~K with accretion heating \citep{2023Marleau}. Thereby, the temperature in Figure~\ref{fig:temperature1} near the planet is only a lower estimate of an accretion-less passively-heated embedded companion.
However, given the size of the companion relative to the host star and the obscuring dust in the disc and envelope around the companion, the amount of flux that the circumstellar disc would receive as a result of the companion's accretion is negligible when compared to the stellar flux.
The companion induced spirals can add heating to the disc \citep{2016Rafikov}, this however would affect not necessarily the same angluar position as the disc, as the shock front trails the planets location in the disc beyond.

\subsection{PDS 70}

\begin{figure}
    \centering
    \includegraphics[width= 1\columnwidth]{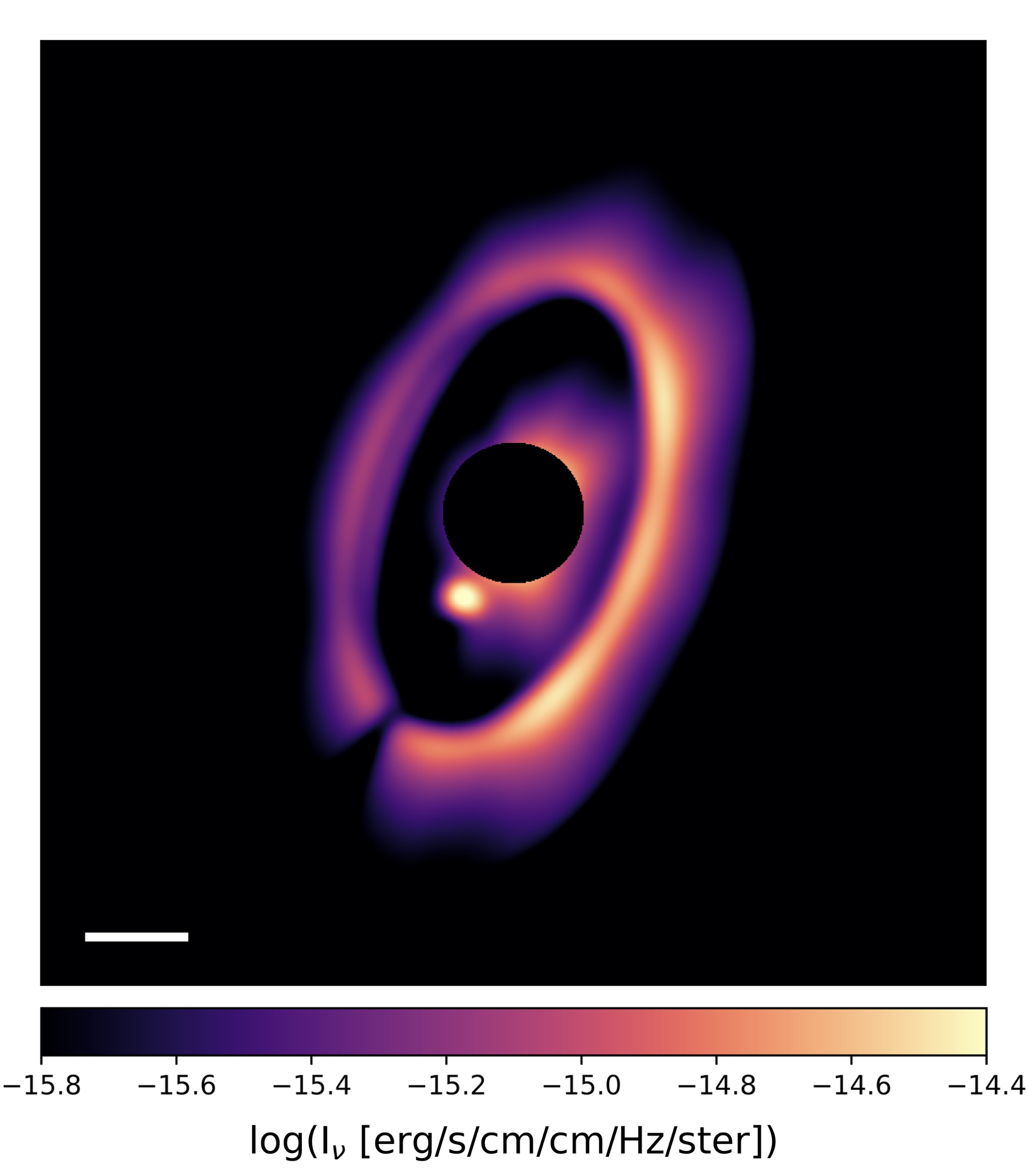}
    \caption{Simulated scattered light image of the PDS 70 system. The white scale bar in the bottom left represents a distance of 20 au.}
    \label{fig:pds70}
\end{figure}

To test the outcome of a shadowing, embedded companion in a disc, we can compare our model to observations of a known system. 
The first planet-system detected within a protoplanetary disc is PDS 70 \citep{2018Keppler, 2019Mesa, 2019Haffert}. Recent multi-band observations by \cite{2024PDS70} highly resolve the scattered light signal of the disc in H, K, J and Y bands. Intensity reduction in the line between planet PDS 70 b and the star is especially prominent in the J and Y bands ($0.9-1.4~\mu $m).

In order to compare this system to the full effect of a planet embedded in a circumplanetary disc up to $1~r_\mathrm{H}$, we run a simulation of the system as shown in Fig.~\ref{fig:pds70}. For this simulation, we adjust the system’s physical parameters based on \cite{2024PDS70}. The host star, with a mass of 0.82 $M_\odot$ and a blackbody temperature of 4000 K, has a 5 $M_\text{J}$ companion orbiting it at a radial distance of 22 au. Observations reveal an inner cavity that extends up to 50 au, in which the dust surface density is estimated to be $\sim$1\% of the outer disc (50 au to 90 au). In light of this, rather than calculating a specific gap width as we do in Equation \ref{eqn:kanagawa}, we adapt the surface density profile as,
\begin{equation}
\Sigma(r) = \Sigma'(r) \times
\begin{cases} 
0.01 + 0.99 \left[1 -  \exp{\left(-\frac{(r - 45~\text{au})}{2~\text{au}}\right)}^6 \right], & r \geq 45~\text{au} \\ 
0.01, & r < 45~\text{au}
\end{cases}
\end{equation}
Here we again use a modified bell-shaped function (see Equation \ref{eqn:mod_surf_dens}) to establish a transition region between the cavity and the rest of the disc over 45–50 au. The unmodified surface density, $\Sigma'(r)$, follows a power-law distribution of the form $\Sigma'(r) \propto r^{-1.1}$. We use the fact that the disc has a scale height of 2.6\% at 50 au, yielding the expression,
\begin{equation}
    H(r) = H_{50} \left( \frac{r}{50~\mathrm{au}} \right)^{1.25},
\end{equation}
where $H_{50} = 0.026 \times 50~\mathrm{au} = 1.3~\mathrm{au}$. Thermal Monte Carlo scattering simulations are conducted for a single dust species of 1 $\mu$m olivine grains under anisotropic scattering conditions. To enhance realism, we compute the polarised scattering from randomly oriented particles. This added effect is particularly important given the disc's inclination of $\sim55\%$.
The disc is imaged at 1.2 $\mu$m, coinciding to the effective wavelength midpoint of the J band. The resulting image is then convolved with a PSF with a FWHM of 0.04, accounting for the system's relative distance of 110 parsecs.

Our simulation is able to reproduce general scattering features seen in observations throughout the disc and also shows the embedded planet casting a narrow shadow onto the disc edge. The shadow obtained with our simplified modelling and convolution approach appears more distinct and pronounced than the reduction seen in observations. This discrepancy may be due to assumptions about the size of the circumplanetary disc and/or the resolution and sensitivity of observations at $50~$au. The simulation, here following the prescription outlined in the paper, can help to get a better estimate on the extent of the circumplanetary material by comparing shadow features. 

In our analysis in Section \ref{sec:results}, we consistently observe shadows cast by companions of mass greater than or equal to 14 $M_J$. This doesn't mean that observable shadows cannot be cast by lower mass objects; on the contrary, our findings only act as a lower limit on what we expect to observe for our chosen disc model. Other factors which we keep fixed within our parameter space can lead to easier observation of shadows. For instance, PDS 70 presents a special case which enables shadow observation for a significantly lower mass companion. Our general model assumes a single, shadow-casting companion to be carving its own gap, whereas PDS 70 hosts an additional planet, PDS 70 c, which is estimated to have a semi-major axis of 34 au \citep{2019Haffert} and contributes to gap formation. This leads to a larger zone of depleted material compared to the prediction of the empirical formula of \cite{kanagawa_2016}, effectively increasing the distance between PDS 70 b and the outer edge of the gap. In our original analysis, we see the impact of scattering light, which leads to the decrease in the extent and depth of shadow. A larger cavity in the case of PDS 70 means that the blocked region of light becomes wider before reaching the outer disc where scattering effects become relevant, as such, the shadow is detectable. A second important point to note is the disc aspect ratio. PSD 70 has a significantly lower aspect ratio than our fiducial flared disc model we use in our parameter space investigation. The ratio of the pressure scale heights at 50 au for PDS 70 and our model (Section \ref{sec:method}) is given by $H_{\mathrm{PDS}} / H_{\mathrm{fiducial}}$ = 0.356. A lower scale height ensures that even smaller objects can sufficiently rise above the surface to cast a detectable shadow in scattered light observations. Due to such factors, our analysis acts as an indicator on shadow detectability under conservative (unfavourable) conditions.

It should also be noted that the circumplanetary material surrounding PDS 70 b is directly irradiated, producing a notable emission signal. Additionally, its semimajor axis of 22 au places it in a well-observable region, making it an ideal candidate for detailed study. In contrast, our analysis explores companions orbiting their host stars as close as 5 au, where resolving them remains challenging even with current polarization techniques designed to suppress stellar light. However, while the companion itself may be difficult to detect, its shadow extends much farther into the disc, creating a spatially broader and more prominent feature in scattered light compared to the localized, point-like emission of the planet.

\section{Conclusions} \label{sec:conclusions}

In this study, we explore the shadow features cast by companions onto protoplanetary discs in scattered light. We use radiative transfer simulations run with \texttt{RADMC-3D} to model the disc surface brightness. 
By examining a range of companion masses and orbital distances, we establish empirical relations for shadow width and depth as functions of companion properties. Our results show that within our chosen disc geometry, companions with masses of $\geq 14~M_\mathrm{J}$ consistently cast detectable shadows; we expect lower mass companions to be able to cast shadows in more favourable disc conditions (i.e. flat discs with less flaring or lower aspect ratios). We find through our parameterisation that shadow width scales inversely with radial distance, $\sigma \propto 1/r$.
This inverse relationship suggests that shadow features observed in scattered light could serve as indicators of companion presence, providing a potential new method for estimating companion properties in protoplanetary discs (such as mass and location). We also found that shadow depth does not monotonically decreases but instead exhibits an initial Gaussian peak within the gap profile followed by an exponential decay in the outer disc, further characterising the shadow's impact on disc structure. 

Furthermore, our analysis of disc temperature variations highlights the influence of companion shadows on disc cooling, particularly within the midplane and around the companion's Hill sphere. Although this model presents an upper limit on the shadow's effect due to its static nature, it lays the groundwork for future dynamic studies to understand the evolving thermal structure of the disc and its implications for disc chemistry and planet formation.

\section*{Acknowledgements}

The authors thank Yvonne Unruh for helpful discussions and feedback on the project.
GB was funded by the European Research Council (ERC) under the European Union’s Horizon 2020 research and innovation programme (Grant agreement No. 853022, PEVAP). 
AP acknowledges support from the Royal Society in the form of a University Research Fellowship and Enhanced Expenses Award and funding from the European Union under the European Union's Horizon Europe Research and Innovation Programme 101124282 (EARLYBIRD). Views and opinions expressed are, however, those of the authors only and do not necessarily reflect those of the European Union or the European Research Council.

\section*{Data Availability}

The data underlying this article will be shared on reasonable request to the corresponding authors.



\bibliographystyle{mnras}
\bibliography{main} 




\appendix

\section{\texttt{RADMC-3D} Outputs}
\label{app:outputs}

\begin{figure*}
    \centering
    \makebox[\textwidth][c]{\includegraphics[width=0.97\textwidth]{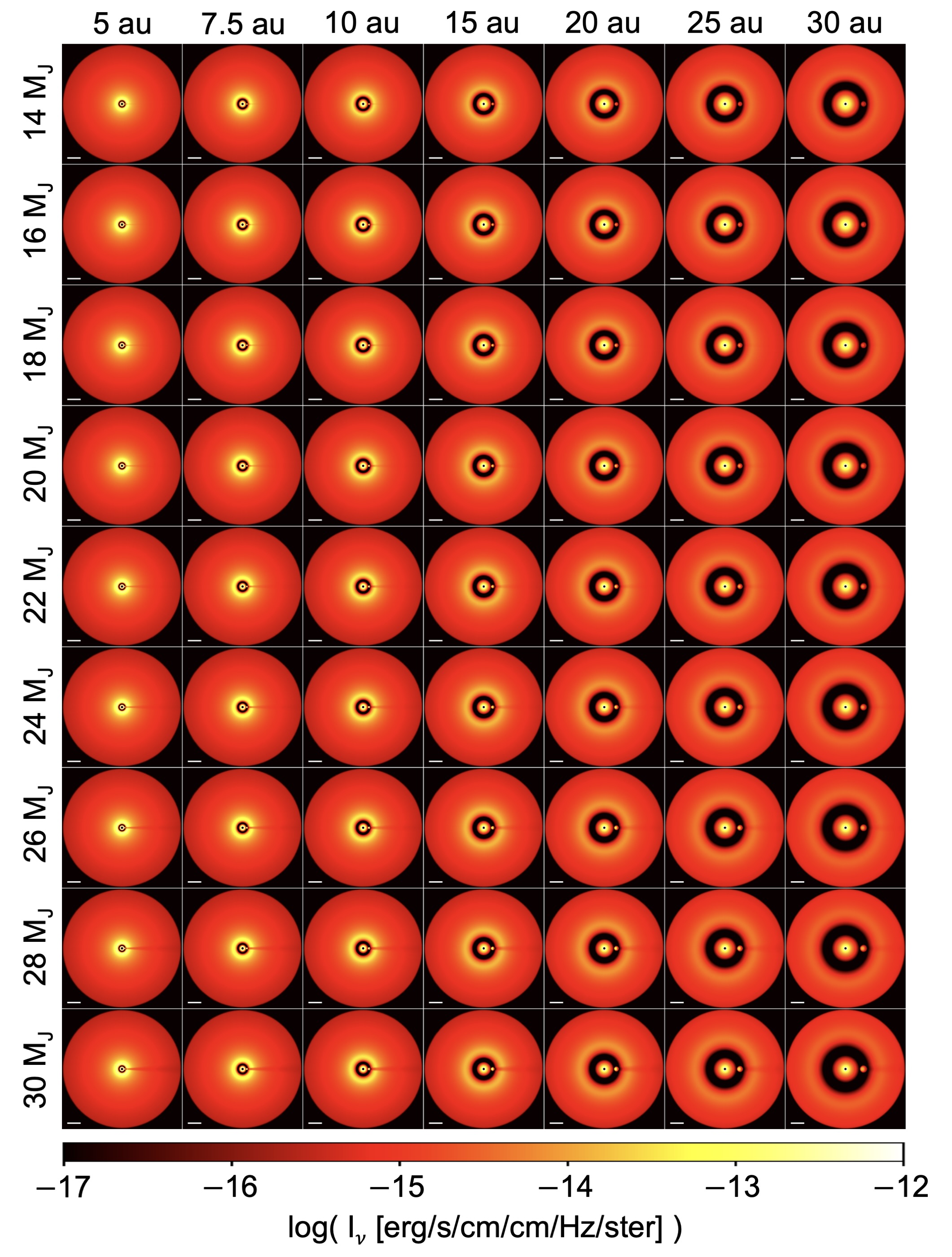}}
    \caption{Synthetic scattered light images, spanning the entire parameter space. Each column denotes the companion's orbital distance while each row indicates companion's mass. The bar in the bottom left corner of each plot scales with 20 au.}
    \label{fig:appendix1}
\end{figure*}

\begin{figure*}
    \centering
    \makebox[\textwidth][c]{\includegraphics[width=0.97\textwidth]{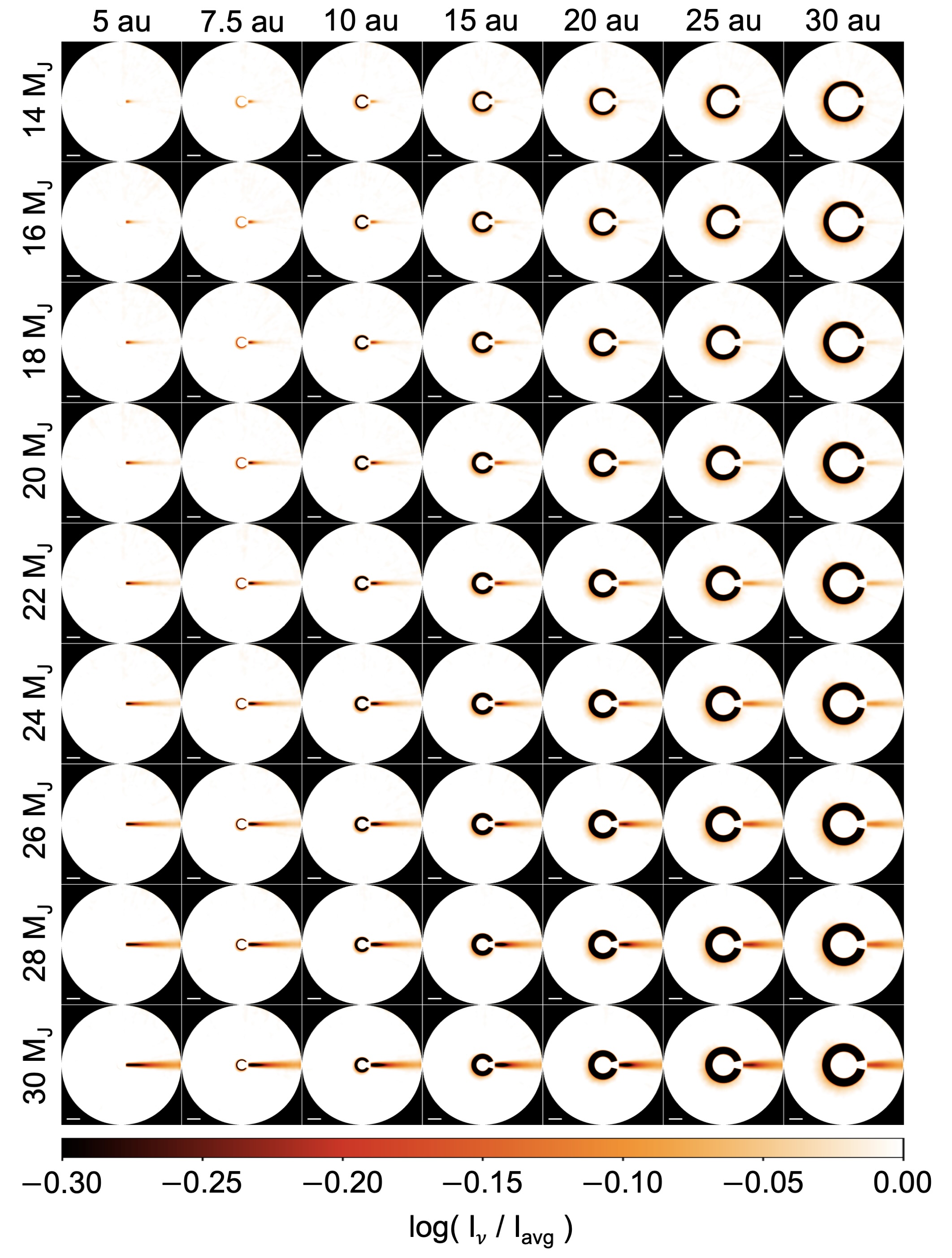}}
    \caption{Residual images, spanning the entire parameter space. Each column denotes the companion's orbital distance while each row indicates companion's mass. The bar in the bottom left corner of each plot scales with 20 au.}
    \label{fig:appendix2}
\end{figure*}

Figure \ref{fig:appendix1} displays the gallery of images obtained as outputs from the radiative transfer simulations. Figure \ref{fig:appendix2} displays the azimuthally normalised residual images for the same set of companion properties.

\section{Geometric Shadow Equation}
\label{sec: geometric}

Our investigation focuses on shadows caused by a 3D Gaussian dust distribution extending out to the Hill sphere of a companion. In this section, we compare it with a lower order, geometric approximation, in which the body is taken to be an optically thick sphere of the size of the Hill radius, and the optical surface of the disc is assumed to be at the pressure scale height, $H(r)$. The shadowed region is defined by a cone with its vertex at the star (origin), and tangential to the Hill sphere. The intersection of a cone with a flat surface (e.g. a disc with no flaring) represents a conic section. In terms of a protoplanetary disc, the amount of flaring of the disc surface affects the shape and extent of this intersection, hence the size of the observable shadow.

\begin{figure}
\centering
\includegraphics[width= 1\columnwidth]{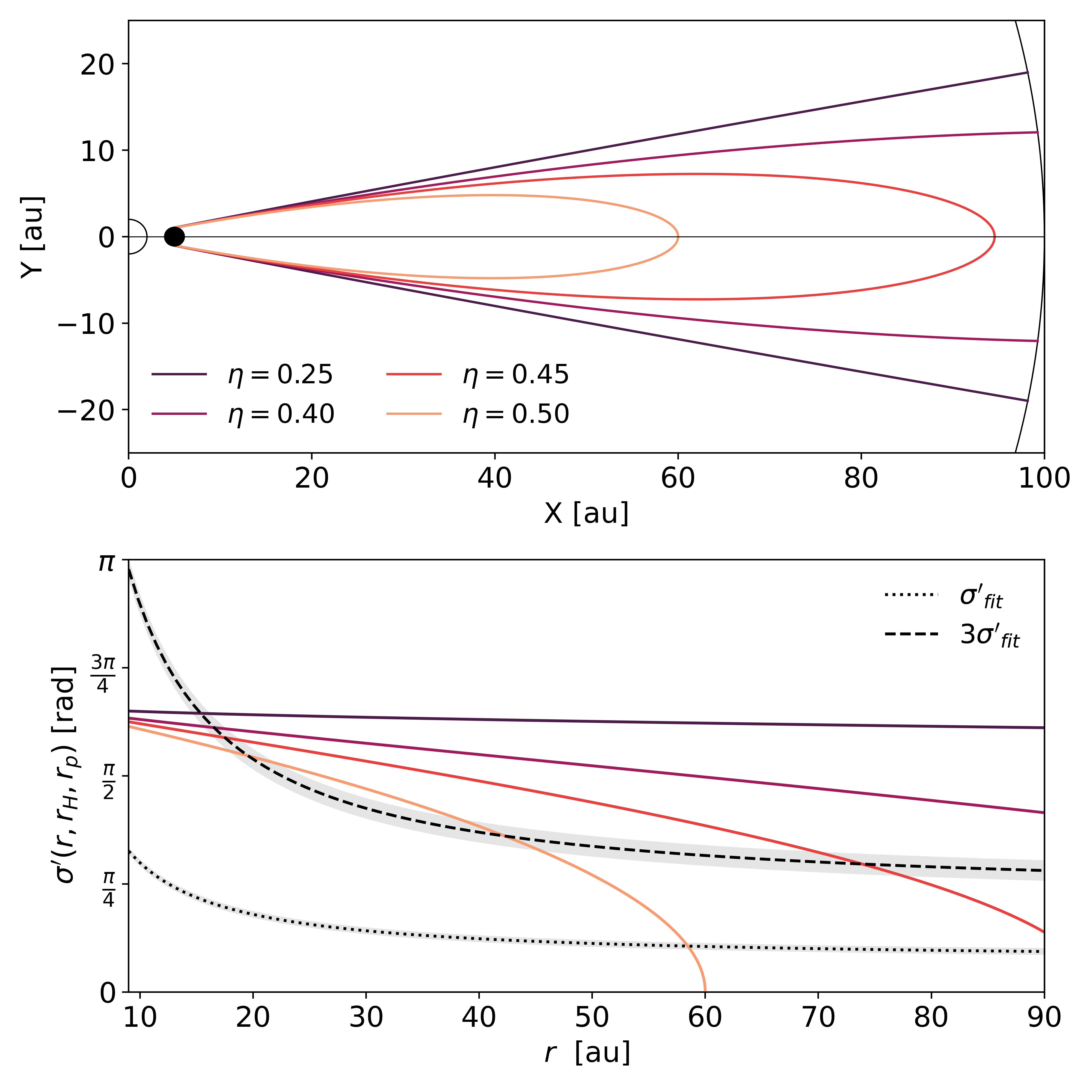}
\caption{\textit{Top:} Top-down view of a disc with a shadow cast by a solid sphere, the size of the Hill radius of a 30~$M_J$ companion orbiting a star at 5~au. The coloured contours represent the shadows cast on the surfaces of discs with different flaring indices, with more flared discs exhibiting smaller shadows. \textit{Bottom:} Scaled shadow width as a function of radial distance. The coloured lines correspond to the shadow geometries shown in the upper panel. The dotted line represents the shadow width (power law) predicted by our analysis considering scattering. The dashed line represents 3 times this width. The grey bands around these lines represent their respective (fitting) uncertainties.}
\label{fig:geometry}
\end{figure}

We first define the equation of the shadow cone. The cone has its vertex at the star and a half-opening angle given by $ \sin \theta = r_H / r_p$, where $r_H$ is the Hill radius and $r_p$ is the orbital radius of the companion. The equation of the cone in cartesian coordinates is hence given by, 

\begin{equation}
    z^2 + y^2 = x^2 \left( \frac{r_H^2}{r_p^2 - r_H^2} \right).
\end{equation}

\noindent The disc optical surface is assumed to be at the pressure scale height, $H(r)$, parameterised as a power law, $z \propto r^{\gamma}$. To determine the shadow boundary, we equate the disc surface to the shadow cone,

\begin{equation}
    \alpha^2 \left( x^2 + y^2 \right)^{\gamma} + y^2 - x^2 \left( \frac{r_H^2}{r_p^2 - r_H^2} \right) = 0,
\end{equation}

\noindent where $\alpha$ is the pressure scale height at $r = 1$ au. This relation can be explicitly formulated in polar coordinates, introducing  $x = r \cos \varphi$ and $y = r \sin \varphi$, the shadow equation transforms into,

\begin{equation}
    \alpha^2 r^{2\gamma} + r^2 \sin^2 \varphi - \beta r^2 \cos^2 \varphi = 0,
\end{equation}

\noindent where $\beta = \frac{r_H^2}{r_p^2 - r_H^2}$. Rearranging for $r$, one can obtain the relation,

\begin{equation}
    r^{2(\gamma-1)} = \frac{\beta \cos^2 \varphi - \sin^2 \varphi}{\alpha^2}.
\end{equation}

\noindent Alternatively, we can solve for the angular extent of the shadow, obtaining the shadow equation,

\begin{equation}
    \varphi = \pm \frac{1}{2} \arccos \left( \frac{2\alpha^2 r^{2\eta} + 1 - \beta}{1 + \beta} \right),
\end{equation}

\noindent where $\eta = \gamma - 1$ is the flaring index of the disc.

This equation defines the boundary of the shadow on the outer disc surface, providing a zeroth-order approximation of its projected shape. Figure \ref{fig:geometry} illustrates the morphology and extent of the shadow cast by a 30 $M_J$ companion at 5 au, on discs with varying flaring indices. The scale height of the discs at 1 au is kept constant, at the same value as used in our fiducial disc model throughout this paper. The upper panel presents a top-down view of the shadow geometry, while the lower panel depicts the radial dependence of the shadow width. The shadow width obtained from our full radiative transfer treatment is overlaid as a dashed line, calculated using Equation \ref{eq:width}. In both the geometric and radiative transfer treatments, shadow widths are scaled according to Equation \ref{scaling}.

With respect to the width calculated from the radiative transfer model, we plot both the shadow width, $\sigma$, and the $3\sigma$ boundary, as the majority of the observable shadowed region is expected to fall within this region. Comparing this $3\sigma$ boundary with the geometric predictions reveals that none of the geometric approximations closely match the radiative transfer results. The radiative transfer model predicts significantly larger shadow widths at smaller radii, near the companion, which can be attributed to complex scattering processes that the geometric model does not account for. In the radiative transfer case, the shadow width exhibits a rapid decay following an approximate 1/r relation, whereas the geometric predictions—based on an arccosine function—do not reproduce this behaviour.

Notably, at larger radii, the geometric model for $\eta = 0.25$, significantly overestimates the shadow width compared to the full scattering treatment. This discrepancy is particularly relevant because $\eta = 0.25$ corresponds to our fiducial disc model, which was used throughout the paper for the full radiative transfer analysis. Beyond a certain radius, the width of the scattering shadow asymptotes and qualitatively resembles the geometric shadow cast on a disc with a flaring index of $\eta = 0.25$, albeit at significantly different values. This suggests that while the overall trend in shadow width at large radii may be comparable between the two methods, the geometric model fails to capture the complexity of radiative transfer effects, as well as the effect of having a 3D Gaussian distribution blocking light as opposed to a solid optically-thick sphere.

Overall, it is evident that scattering plays a crucial role in shaping the $\sigma \propto 1/r$ relation, a feature that cannot be reproduced by lower-order geometric approximations.

\section{Impact of Image Convolution on the Shadow Profile}\label{sec:convolve}

\begin{figure}
\centering
\includegraphics[width= 1\columnwidth]{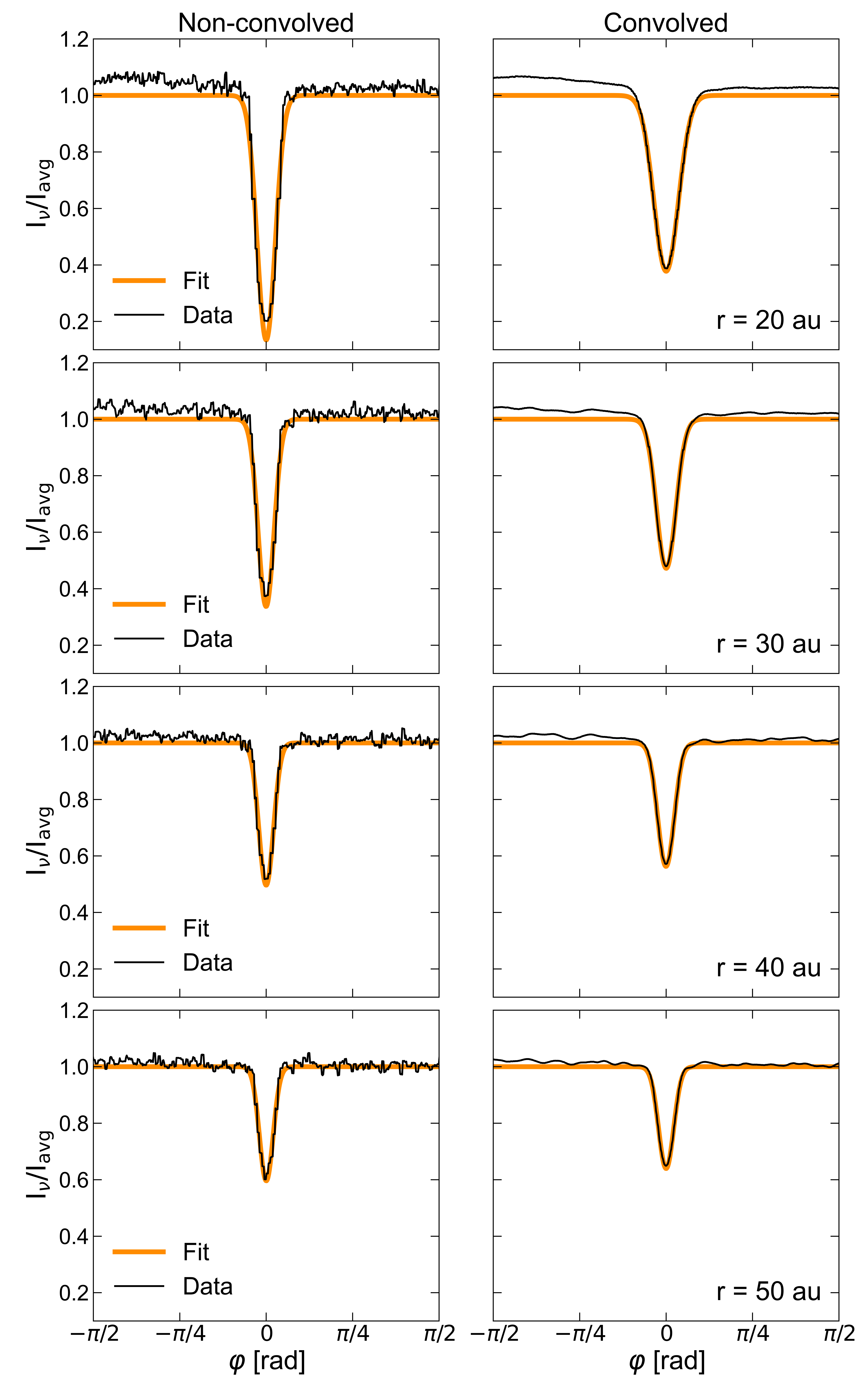}
\caption{Comparison of the azimuthally normalized light intensity profiles for the non-convolved (left column) and convolved (right column) cases at different radial distances. The rows, in descending order, correspond to r = 20 au, 30 au, 40 au, and 50 au. After convolution with the point spread function (PSF), the intensity distribution is smoothed, generally widening the shadow and decreasing its amplitude. This results in a noticeably better Gaussian fit for the convolved case. While the non-convolved distributions fit less well, they can still be approximately modelled as Gaussians.}
\label{fig:conv_comp}
\end{figure}

In this section, we investigate the effect of the convolution step in our data processing, on the ``observed'' shadow properties. To do so, we take the radiative transfer scattered light image output of a single model —specifically, the case of a 30 $M_J$ companion orbiting at 5 au — and apply the full post-processing pipeline as outlined in the main text. We duplicate this process, the only difference between the two cases being whether the image undergoes convolution with a point spread function (PSF) or not. By doing so, we can directly compare the azimuthal dip in light intensity caused by the shadow for both the convolved and non-convolved cases. This comparison is shown in Figure \ref{fig:conv_comp}, where we display the azimuthally normalised light intensity, $\mathrm{I}_{\nu}/\mathrm{I}_\mathrm{avg}$ at different radial distances. Our analysis focuses on two key questions:

\begin{enumerate}
    \item Can a Gaussian function model the shadow profile in the non-convolved case?
    \item If a Gaussian fit is valid, does the radial dependence of the shadow properties differ between the two models?
\end{enumerate}

Beyond the general blurring effect, which reduces noise and small-scale features, convolution primarily results in a broadening of the shadow and a reduction in its contrast (shadow depth). This outcome is consistent with the expected effects of convolving data with a PSF. In general, the non-convolved data exhibits poorer agreement with Gaussian fits, particularly at lower radial distances. At these smaller radii, the fits tend to overshoot, leading to an overestimation of the amplitude relative to the data. This discrepancy is not observed in the convolved case, where the data closely follows a Gaussian profile. Nonetheless, despite these deviations, the non-convolved distributions can still be approximately modelled as Gaussian functions.

\begin{figure}
\centering
\includegraphics[width= 1\columnwidth]{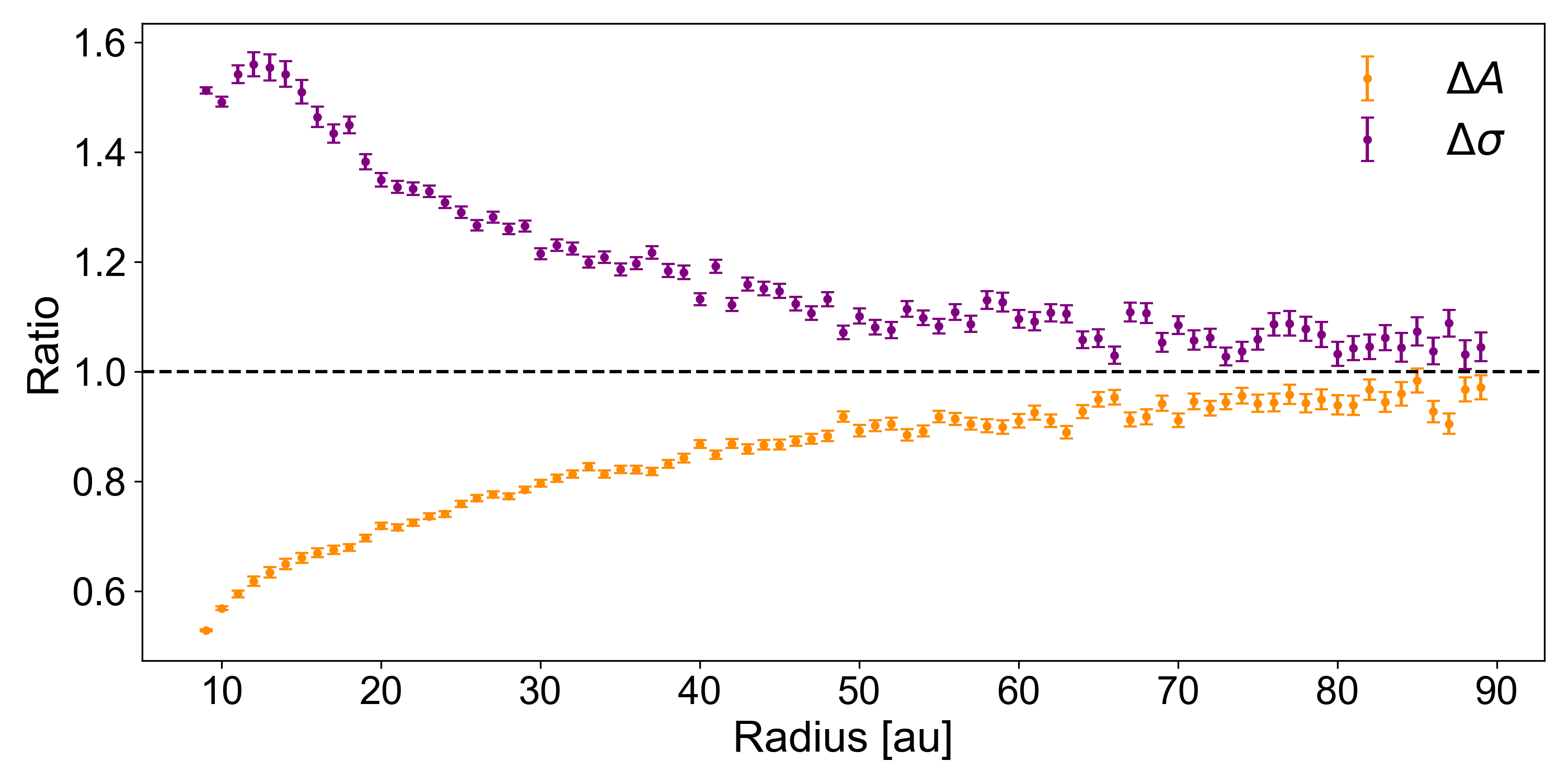}
\caption{Ratios of fitted shadow parameters from non-convolved and convolved signals as a function of radius, for the case of a 30 $M_J$ companion orbiting at 5 au. The amplitude ratio $\Delta A = A / A_c$ is shown in orange, and the width ratio $\Delta \sigma = \sigma / \sigma_c$ is shown in purple. The dashed line indicates unity.}
\label{fig:conv_ratio}
\end{figure}

Given that the shadow can be approximated as a Gaussian in both the non-convolved and convolved signals, we finally look at how the associated parameters (i.e. the shadow width and amplitude) differ between the two cases. This comparison is shown in Figure \ref{fig:conv_ratio}, where we plot the ratios of parameters derived from the non-convolved versus convolved fits, specifically $\Delta \sigma = \sigma/\sigma_c$ and $\Delta A = A/A_c$, where subscript $c$ denotes convolution. It is evident that, at smaller radii, the non-convolved fit overestimates the shadow amplitude while underestimating the width. This effect diminishes with increasing radius, as the ratios gradually approach 1 from either side. Although the convolution step introduces a noticeable difference, we expect real observational data to exhibit similar behaviour due to optical effects arising from the telescope’s geometry and the limitations in angular resolution.

\bsp	
\label{lastpage}
\end{document}